\documentclass[useAMS,usenatbib]{mnras}
\usepackage{journals}
\usepackage{graphicx}
\usepackage{color}
\usepackage{times}
\usepackage{hyperref}
\usepackage{amsmath}
\usepackage{mathtools}

\makeatletter
\DeclareRobustCommand\widecheck[1]{{\mathpalette\@widecheck{#1}}}
\def\@widecheck#1#2{%
    \setbox\z@\hbox{\m@th$#1#2$}%
    \setbox\tw@\hbox{\m@th$#1%
       \widehat{%
          \vrule\@width\z@\@height\ht\z@
          \vrule\@height\z@\@width0.7\wd\z@}$}%
    \dp\tw@-\ht\z@
    \@tempdima\ht\z@ \advance\@tempdima2\ht\tw@ \divide\@tempdima\thr@@
    \setbox\tw@\hbox{%
       \raise\@tempdima\hbox{\scalebox{1}[-1]{\lower\@tempdima\box
\tw@}}}%
    {\ooalign{\box\tw@ \cr \box\z@}}}
\makeatother

\title[Supernovae and their host galaxies -- VI.]{Supernovae and their host galaxies -- VI.
Normal Type~Ia and 91bg-like supernovae in ellipticals}
\author[L.~V.~Barkhudaryan~et~al.]{L.~V.~Barkhudaryan,$^{1}$\thanks{\fontsize{7.8}{9.4}\selectfont{E-mail:
\href{mailto:barkhudaryan@bao.sci.am}{barkhudaryan@bao.sci.am} (LVB);
\href{mailto:hakobyan@bao.sci.am}{hakobyan@bao.sci.am} (AAH)}}
A.~A.~Hakobyan,$^{1}$\textcolor[rgb]{0,0,1}{\footnotemark[1]}
A.~G.~Karapetyan,$^{1}$
\newauthor
G.~A.~Mamon,$^{2}$
D.~Kunth,$^{2}$
V.~Adibekyan$^{3}$
and M.~Turatto$^{4}$
\\
$^{1}$Byurakan Astrophysical Observatory, 0213 Byurakan, Aragatsotn province, Armenia\\
$^{2}$Institut d'Astrophysique de Paris, Sorbonne Universit\'{e}s, UPMC Univ Paris 6 et CNRS, UMR 7095, 98 bis bd Arago, F-75014 Paris, France\\
$^{3}$Instituto de Astrof\'{i}sica e Ci\^{e}ncia do Espa\c{c}o, Universidade do Porto, CAUP, Rua das Estrelas, P-4150-762 Porto, Portugal\\
$^{4}$INAF -- Osservatorio Astronomico di Padova, Vicolo dell'Osservatorio 5, I-35122 Padova, Italy}
\begin{document}

\date{Accepted ---. Received ---; in original form ---}

\pagerange{\pageref{firstpage}--\pageref{lastpage}} \pubyear{2019}

\maketitle

\label{firstpage}

\begin{abstract}
  We present an analysis of
  the galactocentric distributions of the ``normal'' and peculiar ``91bg-like'' subclasses
  of 109 supernovae (SNe) Ia, and study the global parameters of their elliptical hosts.
  The galactocentric distributions of the SN subclasses are consistent with each other,
  and with the radial light distribution of host stellar populations,
  when excluding bias against central SNe.
  Among the global parameters, only the distributions of $u - r$ colours and
  ages are inconsistent significantly between the ellipticals of
  different SN~Ia subclasses: the normal SN hosts are on average bluer/younger
  than those of 91bg-like SNe.
  In the colour--mass diagram, the tail of colour distribution of normal SN hosts stretches
  into the Green Valley -- transitional state of galaxy evolution, while the same tail of
  91bg-like SN hosts barely reaches that region. Therefore,
  the bluer/younger ellipticals might have more residual star formation that gives rise
  to younger ``prompt'' progenitors, resulting in normal SNe~Ia with shorter delay times.
  The redder and older ellipticals that already exhausted their gas for star formation
  may produce significantly less normal SNe with shorter delay times,
  outnumbered by ``delayed'' 91bg-like events.
  The host ages (lower age limit of the delay times) of 91bg-like SNe
  does not extend down to the stellar ages that produce significant $u$-band fluxes
  -- the 91bg-like events have no prompt progenitors.
  Our results favor SN~Ia progenitor models such as He-ignited violent mergers
  that have the potential to explain the observed SN/host properties.
\end{abstract}


\section{Introduction}
\label{intro}

The most energetic and relatively uniform class among Supernovae (SNe) explosions
are Type Ia SNe that were used to discover the accelerating expansion of the Universe
\citep[e.g.][]{1998AJ....116.1009R,1999ApJ...517..565P}.
It is widely accepted that Type Ia SN arises from a thermonuclear explosion of
a carbon-oxygen (CO) white dwarf (WD) in an interacting binary stellar system.
In short, the most favored are single degenerate \citep*[SD;][]{1997Sci...276.1378N}
and double degenerate \citep[DD;][]{1984ApJS...54..335I} progenitor scenarios.
In the SD scenario, a CO~WD accretes material from a main-sequence/subgiant star,
or a red-giant star, or even a helium star, causing the WD mass to reach the
Chandrasekhar mass limit ($\approx 1.4 M_{\odot}$) and explode.
In the DD scenario, a double WD system loses orbital angular momentum due to
gravitational wave emission, leading to coalescence/accretion and explosion.
Recent results suggest that both scenarios are possible
\citep[see e.g.][for a review on various progenitor models]{2016IJMPD..2530024M}.

The fortune of SNe~Ia in cosmology is due to the fact that despite their
moderate inhomogeneity, they are the best standardizable candles in the Universe
thanks to a correlation between their luminosity at maximum light and the shape of
the light-curve (LC), with faster declining objects being fainter
(first proposed by \citealt{1974PhDT.........7R} and \citealt{1977SvA....21..675P}).
This is known as the width-luminosity relation of SN LC \citep{1993ApJ...413L.105P}.
In addition, the LC decline rates $\Delta m_{15}$, i.e. the difference in magnitudes
between the maximum and 15 days after the maximum light,
and colours of SNe~Ia are related: the faster declining LCs correspond to the intrinsically redder
events \citep[e.g.][]{1999AJ....118.1766P,1998AJ....116.1009R}.
The luminosity of a SN~Ia and the $\Delta m_{15}$ depend on the kinetic energy of
the explosion, the mass of radioactive $^{56}{\rm Ni}$ in the ejecta
and opacity \citep[e.g.][]{1982ApJ...253..785A,2007Sci...315..825M}.

Despite the relatively uniform maximum luminosities of Type Ia SNe,
there is increasing evidence for photometric and spectroscopic diversities among them.
In comparison with the observational properties of normal SNe~Ia \citep*[][]{1993AJ....106.2383B},
specialists in the field often refer to two ``traditional'' and most common subclasses of peculiar
SNe~Ia - overluminous ``91T-like'' events with slower declining LCs
\citep[][]{1992ApJ...384L..15F,1992ApJ...387L..33R,1992AJ....103.1632P}
and subluminous ``91bg-like'' SNe with faster declining LCs
\citep[][]{1992AJ....104.1543F,1993AJ....105..301L,1996MNRAS.283....1T}.
These peculiar SNe~Ia make up a considerable fraction of local Type Ia SNe
\citep[e.g.][$\sim 30$ per cent]{2011MNRAS.412.1441L},
and are of crucial importance for understanding SNe~Ia events in general.
A few per cent of other subclasses of peculiar SNe~Ia include the faint but slowly
declining ``02es-like'' SNe, ``02cx-like'' events with low luminosities (also called SNe~Iax),
``Ca-rich'' transients, the extremely luminous so-called ``super-Chandrasekhar'' (also called ``06gz-like'') SNe,
and SNe~Ia showing circumstellar medium interactions
\citep[see e.g.][for a recent review on most of the extremes of Type Ia SNe]{2017hsn..book..317T}.

When considering only the most populated subclasses of SNe~Ia, i.e. normal, 91T- and 91bg-like events,
the lower mass of the host galaxy (the later morphological type or higher the specific star formation rate [SFR]),
the brighter and slower the SNe~Ia that are exploded, on average
\citep[e.g.][]{1996AJ....112.2391H,2001ApJ...554L.193H,2005ApJ...634..210G,2009ApJ...707.1449N,
2011MNRAS.412.1441L,2011ApJ...727..107G,2014ApJ...795..142G}.
91T-like SNe occur in star-forming host galaxies, while such an object has never been discovered
in elliptical galaxies \citep[e.g.][]{2001ApJ...554L.193H,2005ApJ...634..210G,2011MNRAS.412.1441L},
where the stellar population almost always consists of old stars.
91bg-like events prefer host galaxies with elliptical and lenticular morphologies (E--S0),
sometimes they explode also in early-type spirals
\citep[e.g.][]{2001ApJ...554L.193H,2011MNRAS.412.1441L}.
Normal SNe~Ia are discovered in host galaxies with any morphologies
from ellipticals to late-type spirals \citep[e.g.][]{2011MNRAS.412.1441L}.

In the literature, there are many efforts in studying the links between the spectral as well as
LC properties of SNe~Ia and the global as well as local properties at SN explosion sites of
their host galaxies, such as mass, colour, SFR, metallicity, and age of the stellar population
(e.g. \citealt{2000AJ....120.1479H,2000ApJ...542..588I,2005ApJ...634..210G,2008ApJ...685..752G,
2009ApJ...691..661H,2009ApJ...707.1449N,2010MNRAS.406..782S,2011ApJ...740...92G,
2012ApJ...755..125G,2014MNRAS.438.1391P,2015MNRAS.446..354P,
2015MNRAS.448..732A,2016MNRAS.462.1281M,2018ApJ...854...24K};
\citealt*{2019ApJ...874...32R}).
In such studies, SNe~Ia host galaxies with various morphological properties,
e.g. old ellipticals with spherically-distributed stellar content, lenticulars with
an old stellar population in a huge spherical bulge plus a prominent exponential disc,
and spirals with old bulge and young star forming disc components are simultaneously
included in the samples.
In this case, it is difficult to precisely analyse the spatial distribution of SNe,
and associate them with a concrete stellar component (bulge or thick/thin discs,
old or intermediate/young) in the hosts due to different or unknown projection effects
\citep[e.g.][]{2016MNRAS.456.2848H,2017MNRAS.471.1390H}.
In addition, E--S0 and spiral host galaxies have had different evolutionary paths through
major/minor galaxy-galaxy interaction
\citep[e.g.][]{2009MNRAS.394.1713K,2014MNRAS.440..889S,2014MNRAS.442..533M},
and therefore, this important aspect should be clearly distinguished.

In this study, we morphologically select
from the Sloan Digital Sky Survey (SDSS) only elliptical host galaxies of
SNe~Ia, which are known to have the simplest structural properties of the composition in
comparison with lenticular and spiral galaxies \citep[e.g.][]{2009ApJS..182..216K}.
As already mentioned, in these galaxies no 91T-like events have been discovered,
they mostly host normal and 91bg-like SNe \citep[e.g.][]{2001ApJ...554L.193H}.
Therefore, these two subclasses of Type Ia SNe are the subject of study in this
paper.\footnote{The subclasses of Type Ia SNe, discovered in lenticular and spiral
host galaxies, will be the subject of a forthcoming paper in this series.}

Recall that the 91bg-like SNe are unusually red and have peak luminosities that are
$2\pm0.5$ magnitudes lower than do normal SNe~Ia
\citep[the typical peak magnitude of normal SNe~Ia is $M_B\simeq-19.1$~mag,
see][and references therein]{2008MNRAS.385...75T}.
They have faster declining LCs $1.8 \lesssim \Delta m_{15} \lesssim 2.1$,
compared with $\Delta m_{15} \lesssim 1.7$ for normal events,
and their ejecta velocities are small at any epoch in comparison with normal SNe~Ia
\citep[e.g.][]{2005ApJ...623.1011B,2013Sci...340..170W}.
In the post-maximum spectra, particularly notable is the presence of unusually strong
\mbox{O\,{\sc i}\,$\lambda$~7774} and \mbox{Ti\,{\sc ii}} absorption lines.
Despite the recent detection of strong H$\alpha$ in the nebular spectrum of ASASSN-18tb
\citep[a 91bg-like event;][]{2019MNRAS.486.3041K}, there is no evidence that fast declining SNe
are more likely to have late time H$\alpha$ emission \citep[][]{2019ApJ...877L...4S}.
For more details of the spectra and LC properties of 91bg-like events,
the reader is referred to a review by \citet{2017hsn..book..317T}.

The explosion mechanism, which should explain the main characteristics of these events,
including the low $^{56}{\rm Ni}$ masses, is still under debate.
The DD scenario, the helium layer detonation triggered sub-Chandrasekhar mass explosion,
and the scenario of collision of two WDs are competing
\citep[e.g.][]{2000ARA&A..38..191H,2012MNRAS.424.2926M,2013ApJ...770L...8P,
2015MNRAS.454L..61D,2017NatAs...1E.135C}.

In the earlier literature, several attempts have been done to study the projected radial and
surface density distributions of nearby SNe~Ia in morphologically selected elliptical host galaxies
(\citealt*{1980Ap&SS..68..385G,1992A&A...264..428B,2004AstL...30..729T};
\citealt{2008MNRAS.388L..74F}).
These studies showed that, in general, the distribution of Type Ia SNe is consistent
with the light (de~Vaucouleurs) profile of their elliptical host galaxies,
which are dominated by old and metal-rich stellar populations \citep[e.g.][]{2015A&A...581A.103G}.
However, mainly because of the lack of the spectral and LC data, these studies
did not separate the normal and 91bg-like subclasses.
For the first time, \citet[][]{2016AstL...42..495P}
attempted to compare the surface density distributions
of the subclasses of nearby SNe~Ia, in particular for those of the normal and 91bg-like events.
However, the morphological types of SN hosts were not limited to elliptical galaxies only,
thus mixing different progenitor populations from bulges and discs
\citep[see also][]{2008ApJ...685..752G,2019arXiv190410139P}.

On the other hand, \citet[][]{2008ApJ...685..752G}
studied optical absorption-line spectra of 29 early-type
host galaxies of local SNe~Ia and found a higher specific SN rate in E--S0 galaxies
with ages below 3 Gyr than in older hosts.
Recall that the rate of Type Ia SNe
can be represented as a linear combination of ``prompt'' and ``delayed'' (tardy) components
\citep[e.g.][]{2005ApJ...629L..85S}.
The prompt component is more closely related with the recent SFR,
and the delayed component with the total stellar mass of galaxy
\citep[e.g.][]{2005A&A...433..807M,2011MNRAS.412.1473L,2011Ap.....54..301H}.
Therefore, according to \citet{2008ApJ...685..752G}, the higher rate
seen in the youngest E--S0 hosts may be a result of recent star formation and
represents a tail of the prompt SN~Ia progenitors.

Most recently, \citet[][]{2019arXiv190410139P} analysed the explosion sites of eleven
spectroscopically identified nearby 91bg-like SNe in hosts with different morphologies
(including only six E--S0 galaxies) and found that the majority of the stellar populations that
host these events are dominated by old stars with lack of recent star formation evidence.
These authors concluded that the 91bg-like SN progenitors are likely to have delay times,
i.e. the time intervals between the SN~Ia progenitor formation and the subsequent thermonuclear explosion,
much longer \citep[$>6$~Gyr, see also][]{2017NatAs...1E.135C} than the typical delay times of
normal SNe~Ia in star forming environments, whose delay times peak between several hundred Myr and
$\sim 1$~Gyr \citep*[e.g.][]{2014MNRAS.445.1898C,2014ARA&A..52..107M}.

The goal of this paper is to properly address these questions through a comparative
study of the galactocentric distributions of normal and 91bg-like SNe, as well as
through an analysis of the global properties of SNe~Ia hosts
(e.g. stellar mass, metallicity, colour and age of stellar population)
in a well-defined and morphologically non-disturbed sample
of more than 100 relatively nearby elliptical galaxies.

This is the sixth paper of the series following
\citet{2012A&A...544A..81H,2014MNRAS.444.2428H,2016MNRAS.456.2848H,2016MNRAS.459.3130A,2017MNRAS.471.1390H}
and the content is as follows.
The sample selection and reduction are presented in Section~\ref{samplered}.
All the results are presented in Section~\ref{RESults}.
Section~\ref{DISconc} discusses the results with comprehensive interpretations,
and summarizes our conclusions.
To conform to values used in the series of our articles,
Hubble constant $H_0=73 \,\rm km \,s^{-1} \,Mpc^{-1}$ is adopted in this paper.

\section{Sample selection and reduction}
\label{samplered}

We used the updated versions of the Asiago Supernovae
Catalogue\footnote{The ASC was terminated as at 31 December 2017.}
\citep[ASC;][]{1999A&AS..139..531B}
and Open Supernova Catalog \citep[OSC;][]{2017ApJ...835...64G}
to include all spectroscopically classified Type Ia SNe
with distances ${\leq {\rm 200~Mpc}}$
($0.003 \leq z \leq 0.046$),\footnote{Following \citet{2012A&A...544A..81H},
to calculate the luminosity distances of SNe/host galaxies, we used the recession velocities
both corrected to the centroid of the Local Group \citep*{1977ApJ...217..903Y},
and for infall of the Local Group toward Virgo cluster
(\citealt{1998A&A...340...21T}; \citealt*{2002A&A...393...57T}).}
discovered before 9~October 2018.
All SNe are required to have equatorial coordinates and/or
offsets (positions in arcsec) with respect to host galactic nuclei.
We cross-matched these coordinates with the coverage of
the SDSS Data Release Fifteen \citep[DR15;][]{2019ApJS..240...23A}
to identify the host galaxies with elliptical morphology,
using the techniques presented in \citet{2012A&A...544A..81H}.
Many of the identified SNe~Ia host galaxies are already listed in database of
\citet{2012A&A...544A..81H}, which is based on the SDSS DR8.
However, because we added new SNe~Ia, for homogeneity we redid the whole reduction
for the sample of elliptical host galaxies of this study based only on DR15.

Following the approach of \citet[][]{2014MNRAS.444.2428H}, we checked also the levels of
morphological disturbances of the host galaxies using the SDSS images.
Because we are interested in studying the distribution of SNe~Ia in non-disturbed elliptical galaxies,
the hosts with interacting, merging, and post-merging/remnant attributes are removed from the sample.

For three SNe (1980I, 2008gy and 2018ctv), the almost equality of projected distances from
the few nearest elliptical galaxies did not allow to unambiguously assign them to certain hosts.
Therefore, we simply excluded these objects from our sample.

For the remaining SNe~Ia that satisfy the above-mentioned criteria, we carried out an extensive literature search
to collect their spectroscopic subclasses (e.g. normal, 91T-like, 91bg-like and other peculiar events),
which are available at the moment of writing the paper.
To accomplish this, we mainly used the Weizmann Interactive Supernova data REPository \citep[WISeREP;][]{2012PASP..124..668Y},
which is an interactive archive of SN spectra and photometry,
including data of historical events and ongoing surveys/programs.
The archive provides also the important references to the original publications,
which we considered along with the Astronomer's
Telegram\footnote{See \href{http://www.astronomerstelegram.org/}{http://www.astronomerstelegram.org/}.}
(ATEL), website of the Central Bureau for Astronomical
Telegrams\footnote{See \href{http://www.cbat.eps.harvard.edu/iau/cbat.html}{http://www.cbat.eps.harvard.edu/iau/cbat.html}.}
(CBAT) and other supporting publications \citep[e.g.][]{1993AJ....106.2383B,2012MNRAS.425.1789S,2014AN....335..841T}.
In total, we managed to collect the subclasses for 109 SNe in 104 host galaxies:
66 SNe are normal, 41 SNe are 91bg-like, and two SNe are 06gz-like (super-Chandrasekhar) events.
As expected \citep[e.g.][]{2001ApJ...554L.193H,2005ApJ...634..210G},
91T-like events have not been discovered in elliptical galaxies.
On the other hand, less than a dozen of 06gz-like SNe have been discovered so far,
and they have a tendency to explode in low-mass (low-metallicity) late-type galaxies \citep[e.g.][]{2011MNRAS.412.2735T}.
Therefore, they are not the subject of our study, and because of only two such objects in our sample,
further in the article we do not specifically discuss these events and their hosts,
instead we just present them for illustrative purpose.

It is important to note that in this sample of SNe~Ia only seven normal events ($\sim6$ per cent of objects:
1939A, 1957B, 1970J, 1981G, 1982W, 1993ae, and 1993C) were discovered photographically,
while all the other 102 SNe were discovered by visual or mostly CCD searches.

\begin{figure}
\begin{center}$
\begin{array}{@{\hspace{0mm}}c@{\hspace{0mm}}}
\includegraphics[width=1\hsize]{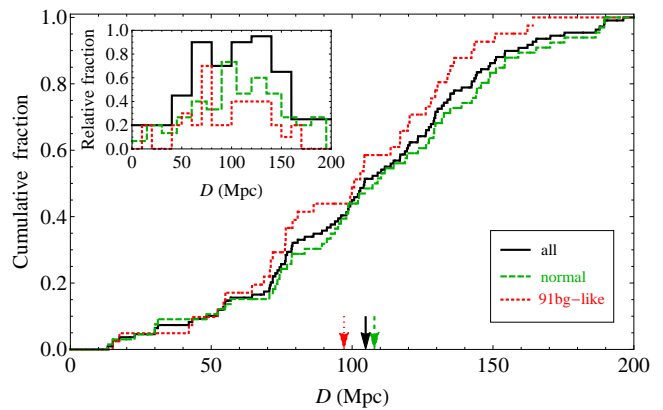}
\end{array}$
\end{center}
\caption{Cumulative and relative (inset) fractions of 109 Type Ia SNe
        (all -- black solid, normal -- green dashed, and 91bg-like -- red dotted)
        in elliptical galaxies as a function of distance.
        The mean values of the distributions are shown by arrows.}
\label{norm91bgdis}
\end{figure}

Fig.~\ref{norm91bgdis} shows the distributions of relative and cumulative fractions
of the subclasses of Type Ia SNe as a function of distances.
As was mentioned, the 91bg-like SNe have peak luminosities that are $\sim2$ magnitudes lower
than do normal SNe~Ia \citep[see][and references therein]{2008MNRAS.385...75T},
therefore the discoveries of 91bg-like events might be complicated at greater distances.
The mean distances of all, normal and 91bg-like SNe are 105, 108 and 97 Mpc
with standard deviations of 43, 44 and 38 Mpc, respectively.
Meanwhile, the two-sample Kolmogorov--Smirnov (KS) and Anderson--Darling (AD) tests\footnote{Traditionally,
we adopted in this article the threshold of 5 per cent for significance levels ($P$-values) of the different tests.
For more details of the statistical tests, the reader is referred to \citet{Engmann+11}.}
showed that the distance distributions of normal and 91bg-like events
are not significantly different ($P_{\rm KS}=0.404$, $P_{\rm AD}=0.238$)
and could thus be drawn from the same parent distribution.
Therefore, our subsamples of normal and 91bg-like SNe and their host galaxies
should not be strongly affected by the redshift-dependent biases
against or in favour of one of the SN subclasses.

We measured the photometry and geometry of the 104 host galaxies
according to the approaches presented in \citet{2012A&A...544A..81H}.
For each host galaxy, we used the fitted $25~{\rm mag~arcsec^{-2}}$ elliptical aperture
in the SDSS $g$-band to obtain the major axis ($D_{25}$),
elongation ($a/b$), and position angle (PA) of the major axis
relative to North in the anticlockwise direction.
The classification of hosts includes also the ratio $10(a-b)/a$: for a projection of
a galaxy with $a$ equal to $b$, the ratio is 0 and the morphological type is E0.
There are only one E5 and ten E4 host galaxies ($\sim10$ per cent of the sample).
The rest of the galaxies are almost evenly distributed in E0--E3 bins.
The mean $D_{25}$ of the hosts is 129 arcsec with the minimum value of 21 arcsec.
The corresponding $u$-, $g$-, $r$-, $i$- and $z$-band fluxes (apparent
magnitudes\footnote{All magnitudes are in the AB system
such that $u_{\rm AB}=u-0.04~{\rm mag}$ and $z_{\rm AB}=z+0.02~{\rm mag}$
($g$, $r$ and $i$ are closer to AB system, see
\href{https://www.sdss.org/dr15/algorithms/fluxcal/}{https://www.sdss.org/dr15/algorithms/fluxcal/}).})
are measured using the $g$-band fitted elliptical aperture.
During the measurements, we masked out bright projected and/or saturated stars.
The apparent/absolute magnitudes and $D_{25}$ values are corrected for
Galactic extinction using the \citet{2011ApJ...737..103S} recalibration of the
\citet*{1998ApJ...500..525S} infrared-based dust map.
These values are not corrected for host galaxy internal extinction because ellipticals
have almost no global extinction, with mean $A_{V}=0.01\pm0.01$~mag \citep[][]{2015A&A...581A.103G}.
Since the redshifts of host galaxies are low ($z \leq 0.046$), the accounted K-corrections
for the magnitudes are mostly negligible and do not exceed 0.2~mag in the $g$-band.
The $D_{25}$ values are also corrected for inclination/elongation effect
according to \citet{1995A&A...296...64B}.

In an elliptical galaxy the real galactocentric distance of SN
can not be calculated using the SN offset from the host galaxy nucleus ($\Delta\alpha$ and $\Delta\delta$).
Instead, we can only calculate the projected galactocentric distance of a SN
($R_{\rm SN}=\sqrt{\Delta\alpha^2+\Delta\delta^2}$),
which is the lower limit of the real galactocentric distance.\footnote{In several cases when SNe offsets
were not available in the above-mentioned catalogues, we calculated $\Delta\alpha$ and $\Delta\delta$
by $\Delta\alpha\approx(\alpha_{\rm SN}-\alpha_{\rm g})\cos\delta_{g}$ and
$\Delta\delta\approx(\delta_{\rm SN}-\delta_{\rm g})$, where $\alpha_{\rm SN}$ and $\delta_{\rm SN}$
are SN coordinates and $\alpha_{\rm g}$ and $\delta_{\rm g}$ are host galaxy coordinates
in equatorial system.}
Following \citet*{2000ApJ...542..588I},
we used the relative projected galactocentric distances ($\widetilde{R}_{\rm SN}=R_{\rm SN}/R_{25}$), i.e.
normalized to $R_{25}=D_{25}/2$ in the $g$-band.

However, for the normalization, the effective radius ($R_{\rm e}$) could be more relevant
being tighter correlated with the stellar surface density distribution
or surface brightness profile of the host galaxy in comparison with
the $R_{25}$ photometric radius \citep[e.g.][]{2009ApJS..182..216K}.
For elliptical galaxies, the surface brightness ($I$) profiles are described by
the S\'{e}rsic law \citep{1963BAAA....6...41S}:
\begin{flalign}
\label{Sersiclow}
& I(R\, | R_{\rm e}) = I_{0} \, {\rm exp} \Bigl\{ -b_{\rm n} \Bigl(\frac{R}{R_{\rm e}}\Bigr)^\frac{1}{n} \Bigr\}\, , &
\end{flalign}
where $R_{\rm e}$ is the radius of a
circle that contains half of the light of the total galaxy (also known as half-light radius),
$I_{0}$ is the central surface brightness of the galaxy, $n$ is the S\'{e}rsic index,
defining the shape of the profile.
An analytical expression that approximates the $b_{\rm n}$ parameter is
${b}_{\rm n} \simeq 1.9992 \, n - 0.3271$ \citep[e.g.][]{1989woga.conf..208C}.
When $n = 4$, the profile, which is called de~Vaucouleurs profile, sufficiently describes
the surface brightness distribution of elliptical galaxies \citep{1948AnAp...11..247D}.
Therefore, following \citet{2008MNRAS.388L..74F}, we also normalized $R_{\rm SN}$ to the $R_{\rm e}$
radii of host galaxies ($\widehat{R}_{\rm SN}=R_{\rm SN}/R_{\rm e}$).

The $g$-band $R_{\rm e}$ radii (in arcsec) of our host galaxies are extracted from the SDSS
where a detailed photometric analysis of galaxies is performed \citep{2001ASPC..238..269L}.
Their pipeline fitted galaxies with a de~Vaucouleurs profile and an exponential profile,\footnote{The
S\'{e}rsic index of $n = 1$ represents the exponential profile of S0--Sm galactic discs \citep{1970ApJ...160..811F}.}
and asked for the linear combination of the two that best-fitted the image,
providing the $R_{\rm e}$ and parameter ${\rm {\texttt fracDeV}}$, which is the fraction of
fluxes contributed from the de~Vaucouleurs profile.
An elliptical galaxy with a pure de~Vaucouleurs profile should have ${\rm {\texttt fracDeV}}=1$,
and a galaxy with pure exponential profile should have ${\rm {\texttt fracDeV}}=0$.
In our morphologically selected sample of hosts, most (about 90 per cent) of the galaxies
have ${\rm {\texttt fracDeV}}>0.8$, where ${\rm {\texttt fracDeV}}=0.8$ roughly corresponds to
S0 galaxies \citep[e.g.][]{2006AJ....131.1288B}.
Only for 14 host galaxies (mostly with $D_{25} > 200$ arcsec),
the SDSS lacks the mentioned model fits or provides unreliable parameters
due to the blending/defragmenting of galaxies with large angular sizes.
For these 14 galaxies, we used our estimations of half-light radii based on the SDSS $g$-band images.

\begin{figure}
\begin{center}$
\begin{array}{@{\hspace{0mm}}c@{\hspace{0mm}}}
\includegraphics[width=1\hsize]{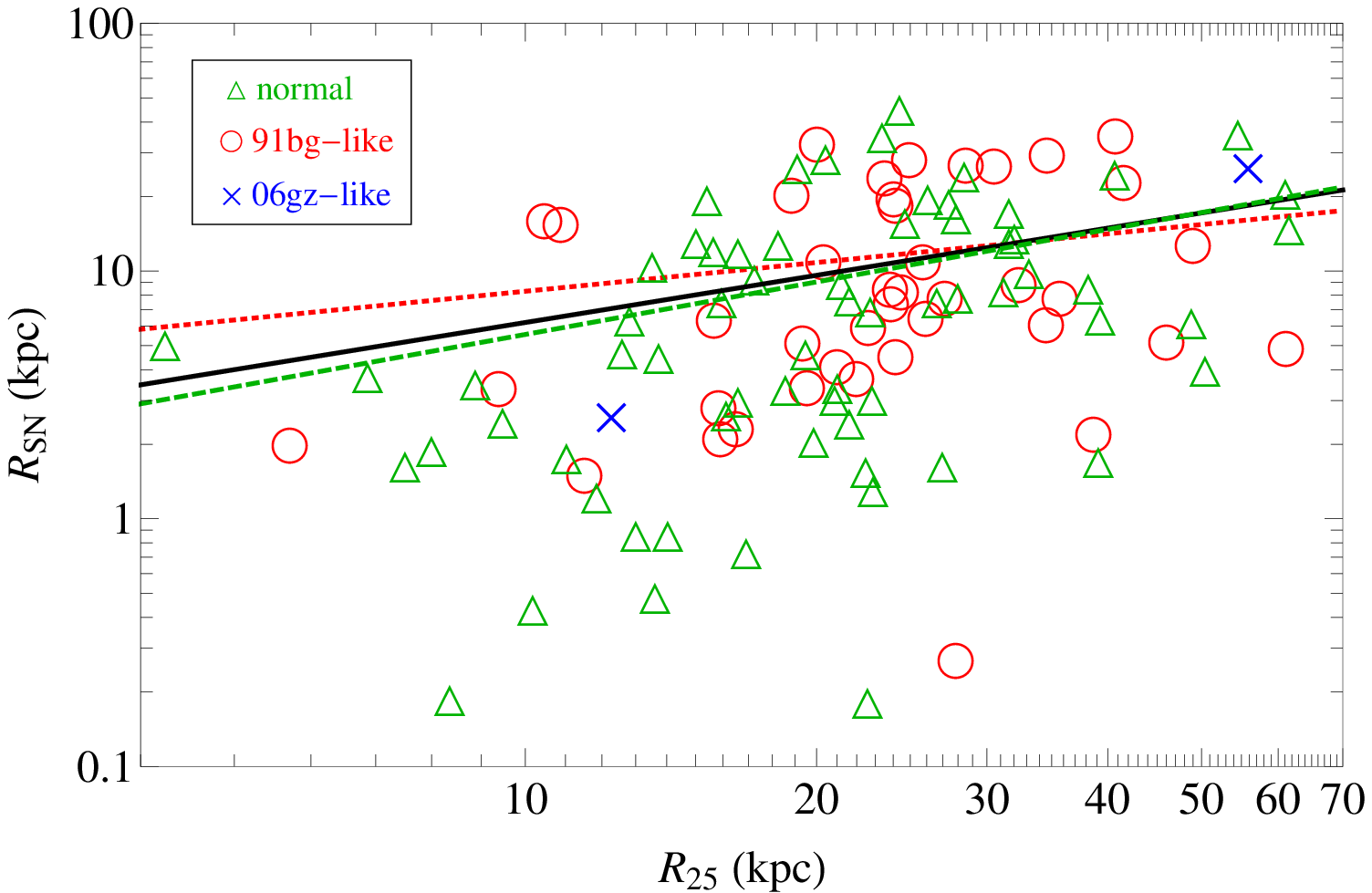}\\
\includegraphics[width=1\hsize]{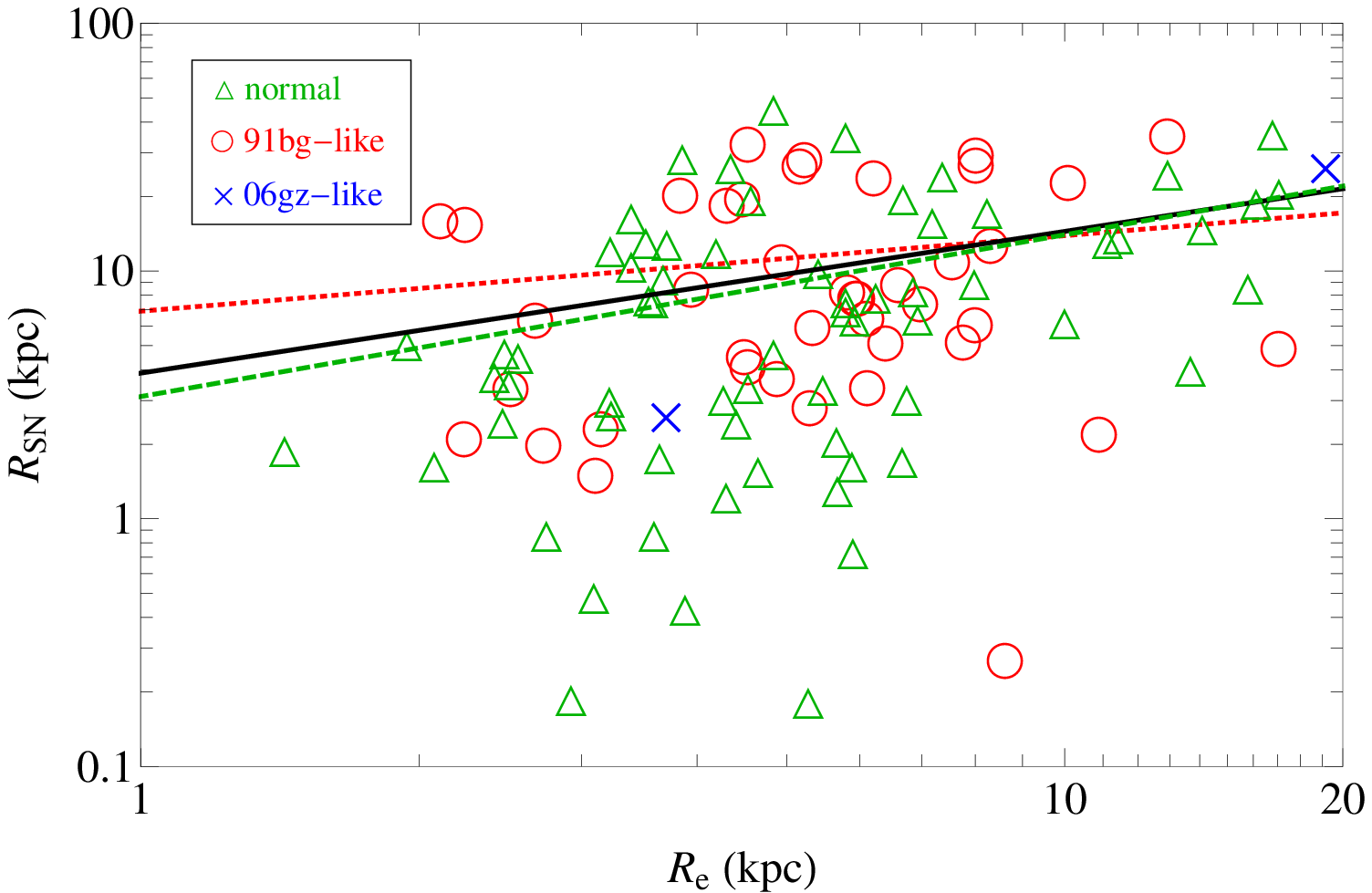}
\end{array}$
\end{center}
\caption{Upper panel: comparison of the projected galactocentric distances
         of SNe~Ia and $R_{25}$ of elliptical host galaxies in kpc.
         Green triangles, red circles and blue crosses show normal,
         91bg-like and 06gz-like SNe, respectively.
         Black solid (all), green dashed (normal) and red dotted (91bg-like) lines
         are best-fits to the samples.
         Bottom panel: same as in upper panel but for $R_{\rm SN}$ versus $R_{\rm e}$.}
\label{RSNR25Rekpc}
\end{figure}
\begin{table}
  \centering
  \begin{minipage}{84mm}
  \caption{The best-fits from Fig.~\ref{RSNR25Rekpc} with results of
           the Spearman's rank correlation test.}
  \tabcolsep 4.4pt
  \label{RSNvsR25eff}
  \begin{tabular}{lrcccc}
  \hline
    \multicolumn{6}{c}{$\log(R_{\rm SN}[{\rm kpc}]) = a + b \, \log(R_{25}[{\rm kpc}])$}\\
    \multicolumn{1}{c}{SN~subclass}&\multicolumn{1}{c}{$N_{\rm SN}$}&
    \multicolumn{1}{c}{$a$}&
    \multicolumn{1}{c}{$b$}&\multicolumn{1}{c}{$r_{\rm s}$}&
    \multicolumn{1}{c}{$P$}\\
  \hline
    all &109&$0.16\pm0.25$&$0.63\pm0.17$&0.457&$\textbf{6} \times \textbf{10}^{-\textbf{7}}$\\
    normal &66&$0.04\pm0.32$&$0.70\pm0.22$&0.481&$\textbf{4} \times \textbf{10}^{-\textbf{5}}$\\
    91bg-like &41&$0.53\pm0.44$&$0.39\pm0.30$&0.316&\textbf{0.044}\\
    \\
    \multicolumn{6}{c}{$\log(R_{\rm SN}[{\rm kpc}]) = a + b \, \log(R_{\rm e}[{\rm kpc}])$}\\
    \\
    all &109&$0.59\pm0.13$&$0.57\pm0.14$&0.364&$\textbf{10}^{-\textbf{4}}$\\
    normal &66&$0.49\pm0.17$&$0.65\pm0.19$&0.414&$\textbf{5} \times \textbf{10}^{-\textbf{4}}$\\
    91bg-like &41&$0.84\pm0.23$&$0.30\pm0.28$&0.162&0.313\\
  \hline
  \end{tabular}
  \parbox{\hsize}{\emph{Notes.} Spearman's coefficient $r_{\rm s}$ is a nonparametric measure of rank correlation
                  ($r_{\rm s}\in[-1;1]$), it assesses how well the relationship between two variables can be described
                  using a monotonic function. The statistically significant correlations
                  ($P$-values $\leq 0.05$) are highlighted in bold.}
\end{minipage}
\end{table}

The $R_{25}$- and $R_{\rm e}$-normalizations are crucial for studying the projected radial distribution of SNe,
because the distribution of linear values of $R_{\rm SN}$
is strongly biased by the greatly different intrinsic sizes of elliptical hosts.
Fig.~\ref{RSNR25Rekpc} illustrates the dependencies of the
$R_{\rm SN}$ on $R_{25}$ and $R_{\rm SN}$ on $R_{\rm e}$ of host galaxies in kpc.
The best-fits from Fig.~\ref{RSNR25Rekpc} and results of
the Spearman's rank correlation test for $R_{\rm SN}$ versus $R_{25}$
and for $R_{\rm SN}$ versus $R_{\rm e}$
(regardless of $\log$ or linear scales) are presented in Table~\ref{RSNvsR25eff}.
The Spearman's rank test indicates significant positive trends ($r_{\rm s}>0$)
between the $R_{\rm SN}$ and $R_{25}$ for all, normal and 91bg-like SNe,
as well as between the $R_{\rm SN}$ and $R_{\rm e}$ for all and normal SNe~Ia.
Only for 91bg-like SNe in the latter case, the trend is positive again but not statistically significant.
In the remainder of this study, we use only normalized projected galactocentric radii of Type Ia SNe,
i.e. $\widetilde{R}_{\rm SN}=R_{\rm SN}/R_{25}$ and $\widehat{R}_{\rm SN}=R_{\rm SN}/R_{\rm e}$.

In addition, we measured the integrated $g$-band flux of
the concentric elliptical aperture, which crosses the position of a SN,
with the same elongation and PA as the host galaxy aperture.
We then normalized this flux to the total flux contained within an elliptical aperture,
retaining the same elongation and PA,
out to distances where the host galaxy flux is consistent with the sky background values.
This fractional radial $g$-band flux is commonly referred as $Fr_g$ and can have values between 0 and 1,
where a value of 0 means that an SN explodes at the center of its host,
while a value of 1 means that the SN explodes at distances where no significant galaxy flux is detected,
i.e. at the edge of the galaxy.
As will be presented in Subsection~\ref{RESults3},
the distribution of $Fr_g$ values allows to compare the radial distribution of SNe~Ia with respect to
that of the $g$-band light of host galaxies, irrespective of their different elongations and S\'{e}rsic indices
\citep[elliptical galaxies can have $n \approx 2$ to 6 in the $g$-band, see e.g.][]{2010ApJS..186..427N,2013MNRAS.435..623V}.
Note that 15 SNe, which are located far outside the elliptical apertures where fluxes are consistent with
the sky background values, are removed from the $Fr_g$ analysis in
Subsection~\ref{RESults3}.\footnote{Their inclusion would artificially increase
the number of SNe~Ia in the fractional radial flux distribution at $Fr_g=1$
(see Subsection~\ref{RESults3}).}
For a complete description of the adopted methodology of $Fr_g$ measurement,
the reader is referred to \citet{2006A&A...453...57J} and \citet{2009MNRAS.399..559A}.

The full database of 109 individual SNe~Ia (SN designation, subclass, source of the subclass,
offset from host galaxy nucleus, and fractional radial $g$-band flux) and their 104 elliptical hosts (galaxy
SDSS designation, distance, $a/b$, PA, $R_{\rm e}$,
corrected $D_{25}$ and $u$-, $g$-, $r$-, $i$-, $z$-band absolute magnitudes)
is available in the online version (Supporting Information) of this article.

\section{Results}
\label{RESults}

With the aim of finding possible links between the properties of SN progenitors and
host stellar populations of elliptical galaxies,
we now study the distributions of projected and normalized galactocentric distances
and fractional radial fluxes of the subclasses of Type Ia SNe (normal and 91bg-like events).
In this section, we also study the possible differences of global properties
(absolute magnitudes, colour, $R_{25}$ and $R_{\rm e}$)
and estimates of the physical parameters (stellar mass, metallicity and age)
of the stellar population of elliptical galaxies
in which the different subclasses of SNe~Ia are discovered.

\subsection{Directional (major vs. minor axes) distributions of SNe~Ia in elliptical host galaxies}
\label{RESults1}

Because the elliptical host galaxies of Type Ia SNe have different elongations
(noted in Section~\ref{samplered}),
it is possible that the distributions of projected galactocentric distances of SNe
along major ($U$) and minor ($V$) axes, normalized to $R_{25}$ or $R_{\rm e}$,
would be different.
Obviously, the projected $U$ and $V$ galactocentric distances (in arcsec) of an SN are
\begin{flalign*}
& U = \Delta\alpha \, \sin{\rm PA} + \Delta\delta \, \cos{\rm PA}\, , &\\
& V = \Delta\alpha \, \cos{\rm PA} - \Delta\delta \, \sin{\rm PA}\, .
\end{flalign*}
Here, as already noted, $\Delta\alpha$ and $\Delta\delta$ are offsets of the SN in equatorial system,
and PA is position angle of the major axis of the elliptical host galaxy.

\begin{table}
  \centering
  \begin{minipage}{84mm}
  \caption{Comparison of the projected and normalized distributions of the subclasses
           of Type Ia SNe along major ($U$) and minor ($V$) axes of elliptical host galaxies.}
  \tabcolsep 4pt
  \label{VRUR_KSAD}
  \begin{tabular}{lrccccc}
  \hline
    \multicolumn{1}{c}{SN~subclass}&\multicolumn{1}{c}{$N_{\rm SN}$}&
    \multicolumn{1}{c}{Subsample~1}&\multicolumn{1}{c}{vs.}&
    \multicolumn{1}{c}{Subsample~2}&\multicolumn{1}{c}{$P_{\rm KS}$}&
    \multicolumn{1}{c}{$P_{\rm AD}$}\\
    &&\multicolumn{1}{c}{$\langle |U|/R_{25} \rangle$}&&
    \multicolumn{1}{c}{$\langle |V|/R_{25} \rangle$}&&\\
  \hline
    all &109&$0.31\pm0.03$& vs. &$0.26\pm0.03$&0.141&0.052\\
    normal &66&$0.27\pm0.03$& vs. &$0.27\pm0.05$&0.438&0.250\\
    91bg-like &41&$0.37\pm0.06$& vs. &$0.26\pm0.05$&0.279&0.143\\
    \\
    &&\multicolumn{1}{c}{$\langle |U|/R_{\rm e} \rangle$}&&
    \multicolumn{1}{c}{$\langle |V|/R_{\rm e} \rangle$}&&\\
    all &109&$1.31\pm0.13$& vs. &$1.13\pm0.15$&0.331&0.099\\
    normal &66&$1.09\pm0.14$& vs. &$1.10\pm0.21$&0.721&0.352\\
    91bg-like &41&$1.68\pm0.27$& vs. &$1.21\pm0.23$&0.420&0.210\\
  \hline
  \end{tabular}
  \parbox{\hsize}{\emph{Notes.} The $P_{\rm KS}$ and $P_{\rm AD}$ are the probabilities from two-sample KS and AD tests,
                  respectively, that the two distributions being compared (with respective mean values) are drawn from the same
                  parent distribution.
                  To calculate the $P_{\rm KS}$ and $P_{\rm AD}$, we used the calibrations by
                  \citet{Massey51} and \citet{Pettitt76}, respectively.}
\end{minipage}
\end{table}
\begin{figure}
\begin{center}$
\begin{array}{@{\hspace{0mm}}c@{\hspace{0mm}}}
\includegraphics[width=1\hsize]{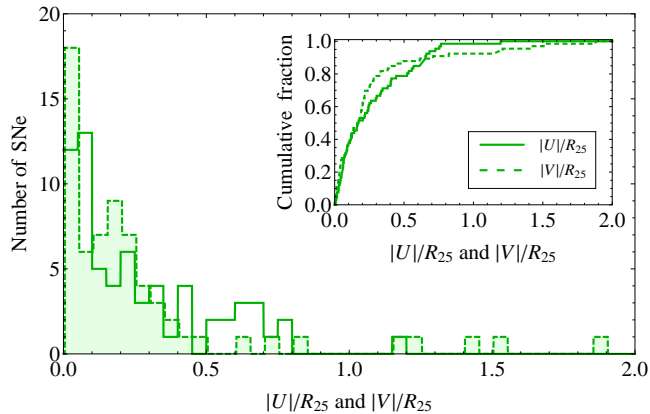}
\end{array}$
\end{center}
\caption{Distributions of $|U|/R_{25}$ (green solid) and $|V|/R_{25}$ (green dashed and filled)
         values for normal SNe~Ia. The inset presents the corresponding cumulative distributions.}
\label{NormHistCum}
\end{figure}

\begin{figure*}
\begin{center}$
\begin{array}{@{\hspace{0mm}}c@{\hspace{0mm}}c@{\hspace{0mm}}}
\includegraphics[width=0.48\hsize]{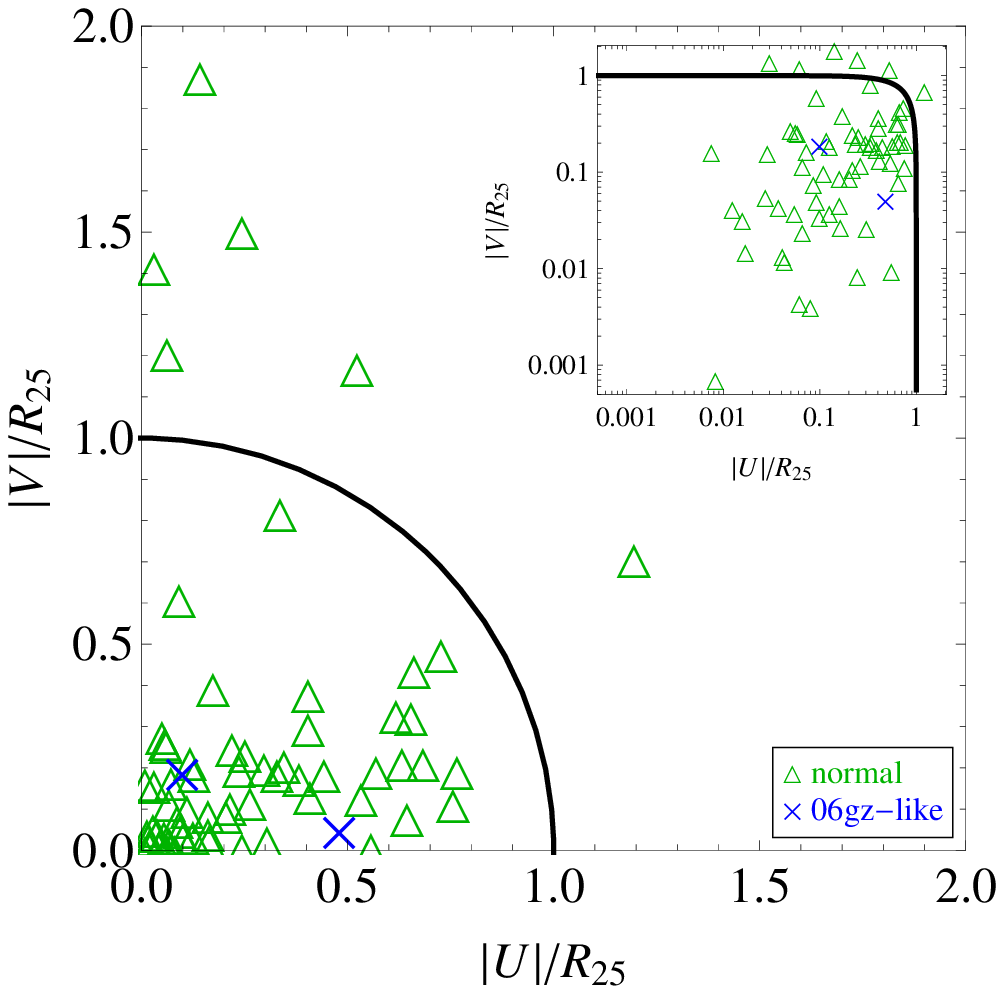} &
\includegraphics[width=0.48\hsize]{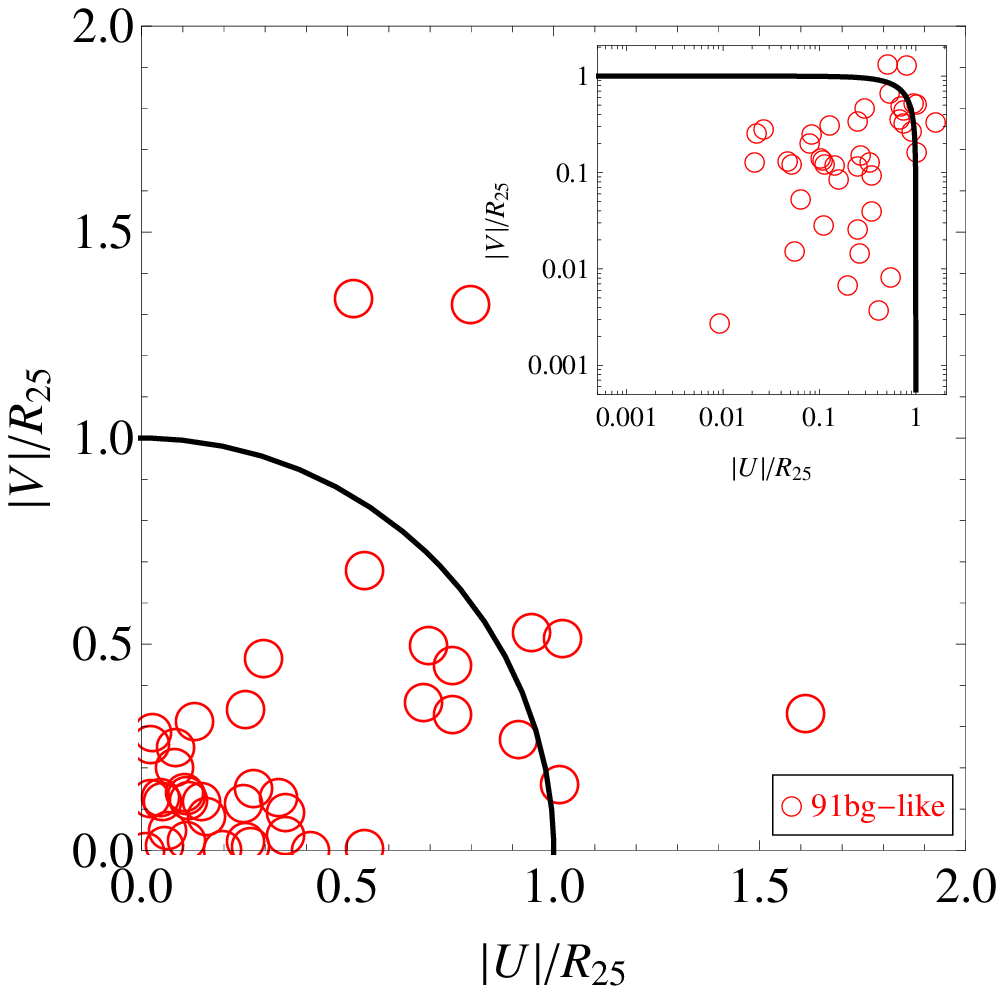} \\
\includegraphics[width=0.48\hsize]{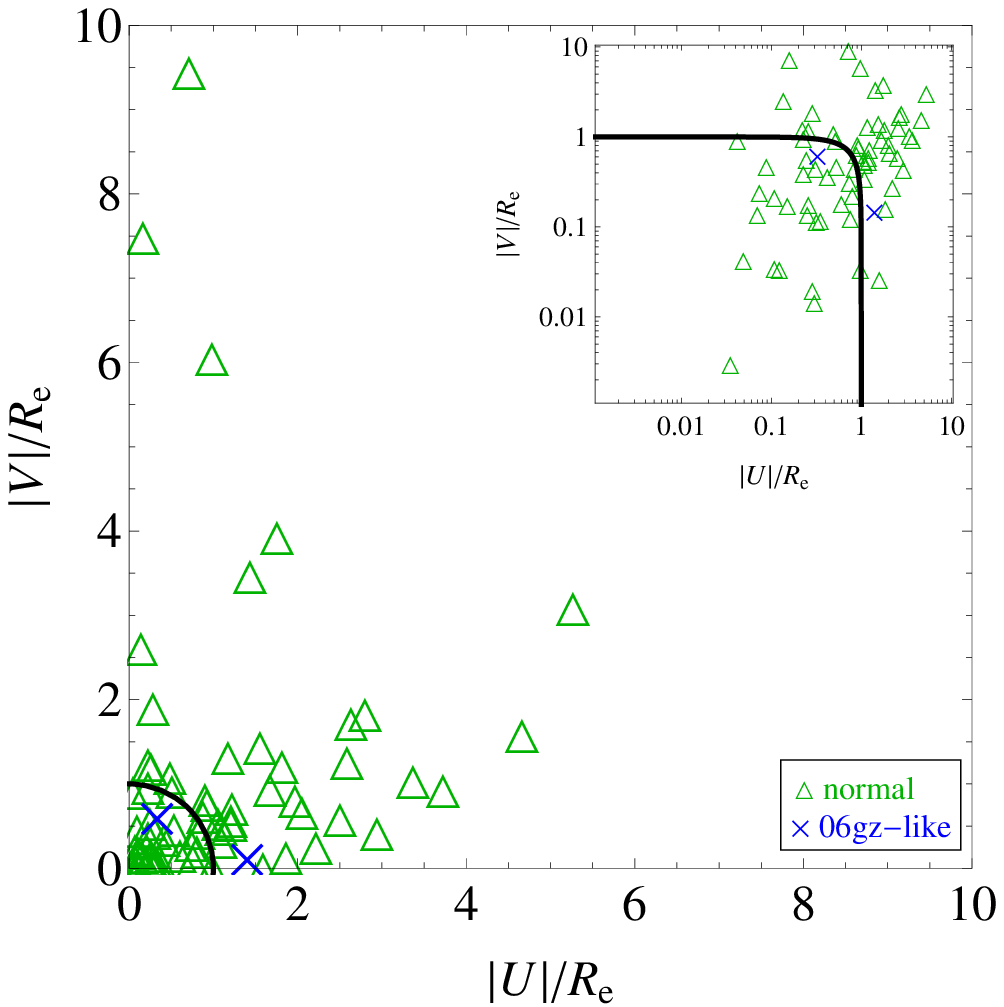} &
\includegraphics[width=0.48\hsize]{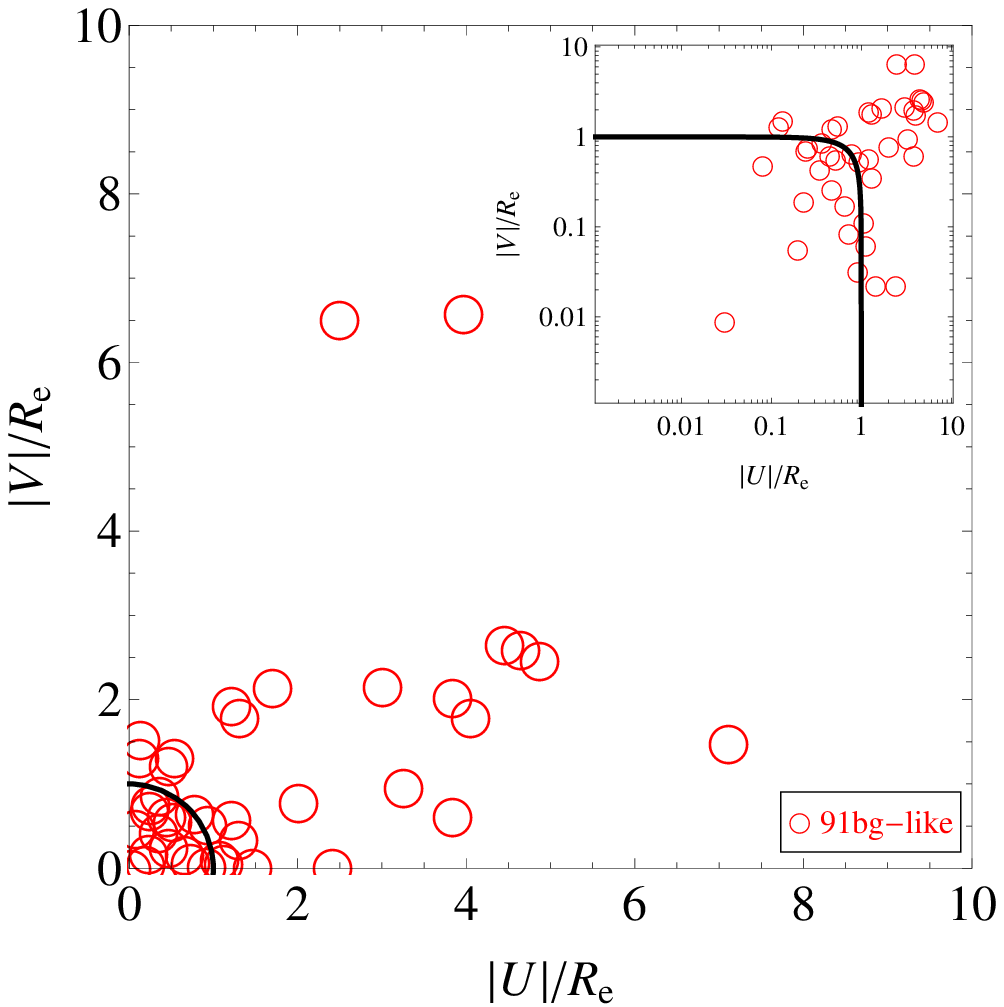}
\end{array}$
\end{center}
\caption{Upper panels: projected and $R_{25}$-normalized distributions of
         the different subclasses of Type Ia SNe along major ($U$) and minor ($V$) axes of elliptical host galaxies.
         The green triangles, blue crosses and red circles represent normal, 06gz-like and 91bg-like SNe, respectively.
         The quarters of big black circles are the host galaxy $R_{25}$ sizes.
         Bottom panels: same as in upper panels but for $R_{\rm e}$ normalization.
         The quarter circles now represent $R_{\rm e}$.
         In the insets, we show the same distributions with $\log$ axes.}
\label{UVR25effSNtype}
\end{figure*}

In the mentioned context, using the two-sample KS and AD tests,
we compare the distributions of $|U|/R_{25}$ versus $|V|/R_{25}$,
as well as the distributions of $|U|/R_{\rm e}$ versus $|V|/R_{\rm e}$
for all, normal and 91bg-like SNe.
Here, the absolute values of $U$ and $V$ are used to increase the statistical power of the tests.
The values of $P_{\rm KS}$ and $P_{\rm AD}$ in Table~\ref{VRUR_KSAD} show that
the distributions of projected and normalized galactocentric distances of SNe
along major and minor axes are consistent between each other.
Only the $P_{\rm AD}$ values for the entire sample of SNe~Ia are close to the rejection threshold of 0.05,
however when we split the sample between normal and 91bg-like events,
both the $P$-values of KS and AD tests become clearly above the threshold.
Therefore, the different elongations of elliptical host galaxies in our sample have negligible impact,
if any, on the sky plane projection of the spherical 3D distribution of SNe~Ia.
For illustration, in Fig.~\ref{NormHistCum} we show the histograms and
cumulative distributions of $|U|/R_{25}$ and $|V|/R_{25}$ for normal SNe~Ia.
Comparison of the same distributions for 91bg-like SNe looks similar
(also for the cases with $R_{\rm e}$ normalization).
Fig.~\ref{UVR25effSNtype} shows the projected distributions of the subclasses of Type Ia SNe
with $R_{25}$ and $R_{\rm e}$ normalizations.

\subsection{The radial distributions of SNe~Ia in ellipticals}
\label{RESults2}

As already mentioned above, the light profiles of elliptical galaxies are characterized with a continuous distribution
according to the \citeauthor{1963BAAA....6...41S} law with mean index $n\approx4$,
when considering the families of ellipticals from dwarfs to giants
\citep[for the $g$-band see e.g.][]{2010ApJS..186..427N,2013MNRAS.435..623V}.
In addition, \citet{2008MNRAS.388L..74F} have already shown that the projected surface density distribution of
Type Ia SNe in morphologically selected early-type host galaxies is consistent with the de~Vaucouleurs profile ($n=4$)
in the $0.2 < R_{\rm SN}/R_{\rm e} < 4$ radial range \citep[see also][]{2004AstL...30..729T,2010ApJ...715.1021D}.
However, in the literature a comprehensive analysis of the surface density distributions of normal and 91bg-like SNe
in well-defined elliptical host galaxies with different radius normalizations
has not yet been performed and, as already mentioned, is one of the main goals of the present study.

\begin{table}
  \centering
  \begin{minipage}{84mm}
  \caption{Consistency of the distribution of projected and normalized galactocentric distances of SNe~Ia with
           the surface density model of S\'ersic profile with $n=4$ (de~Vaucouleurs profile) in elliptical host galaxies.}
  \tabcolsep 6.8pt
  \label{RSNR25Re_deV_KSAD}
  \begin{tabular}{lrrccc}
  \hline
    \multicolumn{1}{c}{SN~subclass}&\multicolumn{1}{c}{$\widetilde{R}_{\rm SN}\geq$}&
    \multicolumn{1}{c}{$N_{\rm SN}$}&
    \multicolumn{1}{c}{$\widetilde{R}_{\rm e}^{\rm SN}$}&\multicolumn{1}{c}{$P_{\rm KS}$}&
    \multicolumn{1}{c}{$P_{\rm AD}$}\\
  \hline
    all &0&109&$0.28\pm0.03$&\textbf{0.008}&\textbf{0.014}\\
    normal &0&66&$0.26\pm0.04$&0.184&0.129\\
    91bg-like &0&41&$0.31\pm0.04$&\textbf{0.036}&\textbf{0.041}\\
  \\
    all &0.1&94&$0.18\pm0.02$&0.231&0.093\\
    normal &0.1&54&$0.18\pm0.02$&0.369&0.222\\
    91bg-like &0.1&38&$0.18\pm0.04$&0.263&0.185\\
  \\
    &\multicolumn{1}{c}{$\widehat{R}_{\rm SN}\geq$}&&\multicolumn{1}{c}{$\widehat{R}_{\rm e}^{\rm SN}$}&&\\
    all &0&109&$1.15\pm0.12$&\textbf{0.016}&\textbf{0.034}\\
    normal &0&66&$1.03\pm0.09$&0.176&0.212\\
    91bg-like &0&41&$1.38\pm0.15$&0.059&0.068\\
  \\
    all &0.4&92&$0.80\pm0.12$&0.132&0.093\\
    normal &0.4&52&$0.78\pm0.18$&0.144&0.196\\
    91bg-like &0.4&38&$0.87\pm0.21$&0.514&0.206\\
  \hline
  \end{tabular}
  \parbox{\hsize}{\emph{Notes.} The $P_{\rm KS}$ and $P_{\rm AD}$ are the probabilities from
                  one-sample KS and AD tests, respectively, that the distributions of SNe~Ia are drawn
                  from the best-fitting de~Vaucouleurs surface density profiles with the maximum likelihood
                  values of $\widetilde{R}_{\rm e}^{\rm SN}=R_{\rm e}^{\rm SN}/R_{25}$ and
                  $\widehat{R}_{\rm e}^{\rm SN}=R_{\rm e}^{\rm SN}/R_{\rm e}$ (with bootstrapped errors, repeated $10^3$ times).
                  The $P_{\rm KS}$ and $P_{\rm AD}$ are calculated using the calibrations by
                  \citet{Massey51} and \citet{1986gft..book.....D}, respectively.
                  The statistically significant deviations from de~Vaucouleurs profile
                  ($P$-values $\leq 0.05$) are highlighted in bold.}
\end{minipage}
\end{table}

Using maximum likelihood estimation (MLE) method, we fit the distribution of projected and
$R_{25}$-normalized galactocentric radii of Type Ia SNe ($\widetilde{R}_{\rm SN}=R_{\rm SN}/R_{25}$)
with the surface density model of S\'ersic profile with $n=4$.
If the surface density ($\Sigma$) of SNe~Ia is described by
a S\'ersic function of $\widetilde{R}_{\rm SN}$ (see Eq.~[\ref{Sersiclow}]),
then the probability that a SN is observed at $\widetilde{R}_{\rm SN}$ radius, i.e. probability density function (PDF), is
\begin{flalign}
\label{pdfIa}
& p(\widetilde{R}_{\rm SN} | \widetilde{R}_{\rm e}^{\rm SN}) =
 \dfrac{\widetilde{R}_{\rm SN} \, \Sigma(\widetilde{R}_{\rm SN} | \widetilde{R}_{\rm e}^{\rm SN})}
 {\int_{0}^\infty \widetilde{R}_{\rm SN} \, \Sigma(\widetilde{R}_{\rm SN} | \widetilde{R}_{\rm e}^{\rm SN}) \, {\rm d}\widetilde{R}_{\rm SN}}\, ,&
\end{flalign}
where $\widetilde{R}_{\rm e}^{\rm SN}=R_{\rm e}^{\rm SN}/R_{25}$ is normalized effective radius of SN distribution.
The likelihood of the set of $\{\widetilde{R}_{{\rm SN}\, i}\}$ is
\begin{flalign}
& {\cal L}(\widetilde{R}_{\rm e}^{\rm SN}) = \prod_{i=1}^{N_{\rm SN}} p(\widetilde{R}_{{\rm SN}\, i} | \widetilde{R}_{\rm e}^{\rm SN})\, ,&
\end{flalign}
and thus maximizing $\ln({\cal L})$ we get the effective radii of SN distributions
for the subclasses of Type Ia SNe.

\begin{figure*}
\begin{center}$
\begin{array}{@{\hspace{0mm}}c@{\hspace{0mm}}c@{\hspace{0mm}}}
\includegraphics[width=0.485\hsize]{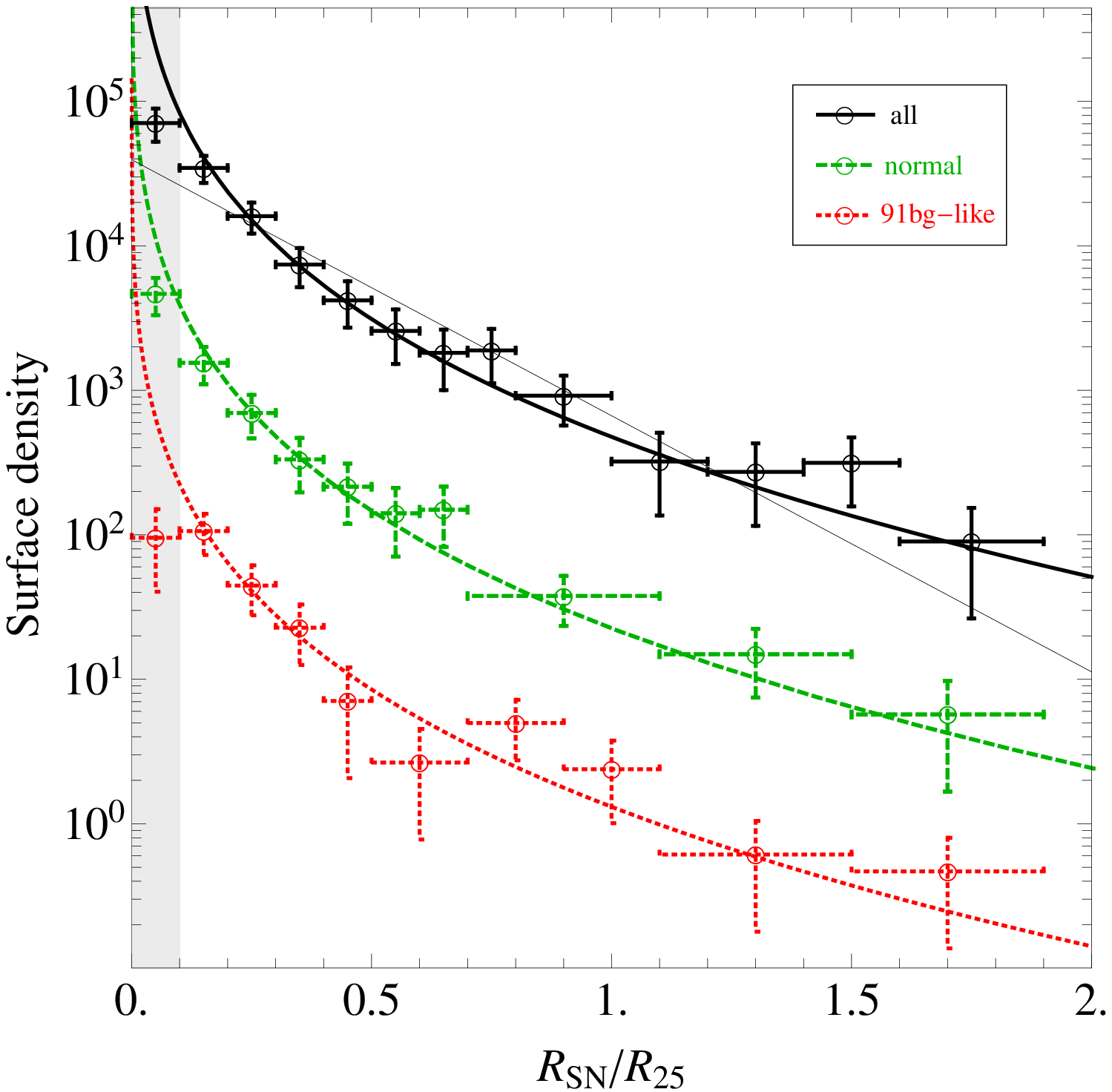} &
\includegraphics[width=0.5\hsize]{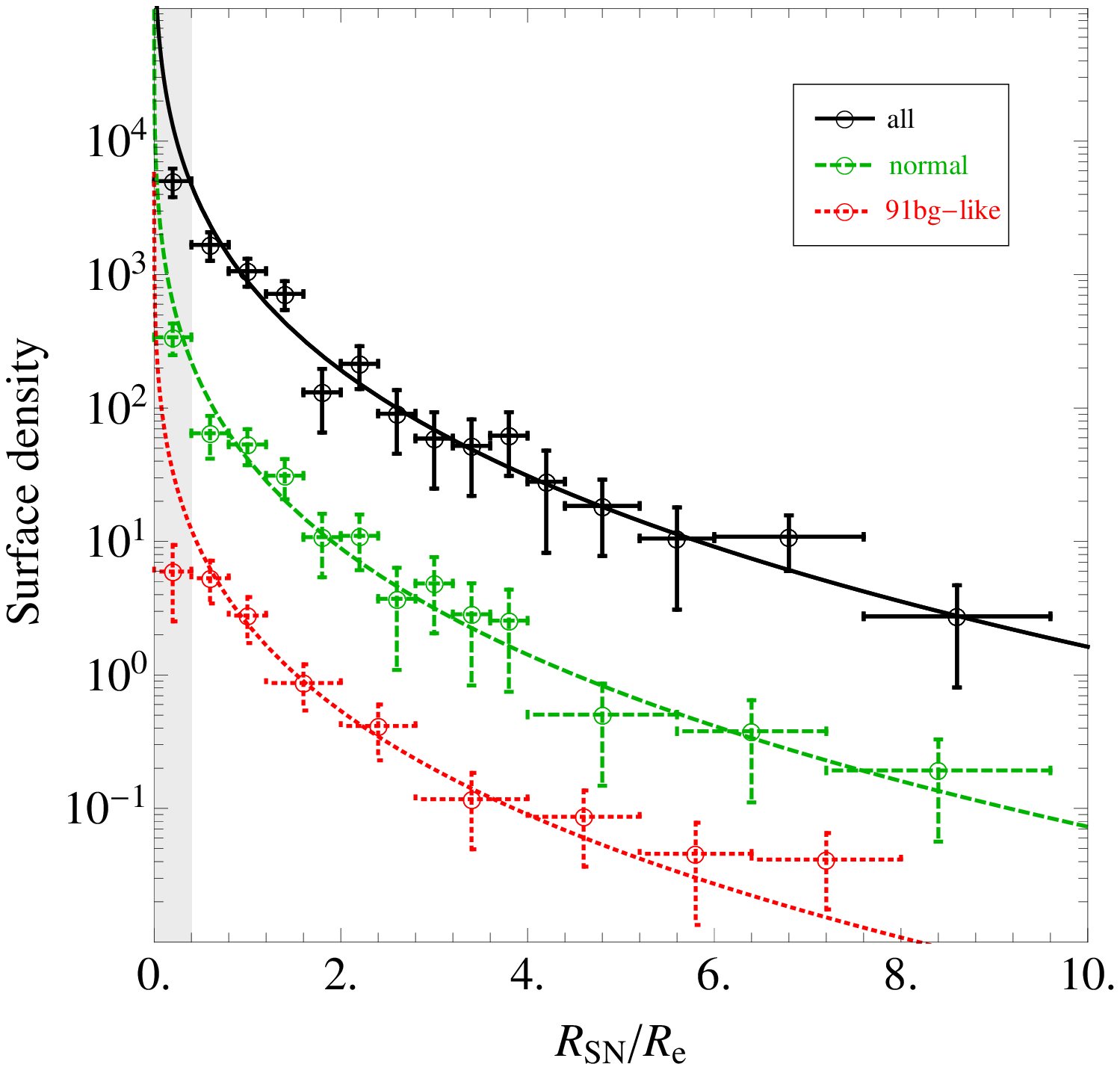}
\end{array}$
\end{center}
\caption{Left: $R_{25}$-normalized surface density distributions (arbitrary scaled) of all (black solid),
         normal (green dashed) and 91bg-like (red dotted) SNe~Ia in elliptical host galaxies.
         The vertical error bars assume a Poisson distribution.
         The horizontal bars show the bin sizes that are increased at the edges of galaxies to include at least two SNe in each.
         The different curves show the maximum likelihood de~Vaucouleurs surface density profiles,
         estimated using the inner-truncated distributions (outside the shaded area).
         For all SNe, the best-fitting inner-truncated exponential profile (black thin line) is also shown.
         For better visibility, the distributions with their best-fitting profiles are shifted vertically
         (to avoid falling one onto another).
         Right: same as in left panel but for $R_{\rm e}$ normalization.}
\label{SD25eff}
\end{figure*}

At the same time,
to check whether the distributions of SNe~Ia follow the best-fit de~Vaucouleurs profiles,
we perform one-sample KS and AD tests on the cumulative distributions of
the projected and normalized galactocentric distances of SNe.
In general, the cumulative distribution function (CDF) of S\'ersic model can be expressed as the integral of
its PDF (see Eq.~[\ref{pdfIa}]) as follows:
\begin{flalign}
E(\widetilde{R}_{\rm SN}) &=  \int_{-\infty}^{\widetilde{R}_{\rm SN}} p(t \, | \widetilde{R}_{\rm e}^{\rm SN}) \, {\rm d}t&\notag\\
&= 1 - \Gamma\Bigl(2n,\, b_{\rm n}\Bigl(\frac{\widetilde{R}_{\rm SN}}{\widetilde{R}_{\rm e}^{\rm SN}}\Bigr)^\frac{1}{n}\Bigr)\Big/\Gamma\Bigl(2n\Bigr)\, , \label{cdfIa}
\end{flalign}
where
\begin{flalign*}
& \Gamma(z) = \int_{0}^\infty t^{z-1} e^{-t} d{t}&
\end{flalign*}
and
\begin{flalign*}
& \Gamma(a,\, z) = \int_{z}^\infty t^{a-1} e^{-t} d{t}&
\end{flalign*}
are the complete and upper incomplete gamma functions, respectively.

For the $R_{\rm e}$ normalization, in the above-mentioned formulae (Eqs.~[\ref{pdfIa}--\ref{cdfIa}])
we simply replace $\widetilde{R}_{\rm SN}$ with $\widehat{R}_{\rm SN}=R_{\rm SN}/R_{\rm e}$
and therefore $\widetilde{R}_{\rm e}^{\rm SN}$ with $\widehat{R}_{\rm e}^{\rm SN}=R_{\rm e}^{\rm SN}/R_{\rm e}$.
The estimated $\widetilde{R}_{\rm e}^{\rm SN}$ and $\widehat{R}_{\rm e}^{\rm SN}$ effective radii, and
the $P_{\rm KS}$ and $P_{\rm AD}$ probabilities that the distributions of SNe~Ia are drawn
from the best-fitting de~Vaucouleurs surface density profiles
(S\'ersic model with $n=4$) are listed in Table~\ref{RSNR25Re_deV_KSAD}.

From the $P$-values in Table~\ref{RSNR25Re_deV_KSAD},
we see that the global ($\widetilde{R}_{\rm SN} \geq 0$ and $\widehat{R}_{\rm SN} \geq 0$)
surface density distributions of Type Ia SNe in elliptical host galaxies are
not consistent with the de~Vaucouleurs profiles.
When splitting the sample between the subclasses of SNe~Ia,
we see that the significant inconsistency exists for the $\widetilde{R}_{\rm SN}$ distribution of 91bg-like events,
and the marginal inconsistency takes place for the $\widehat{R}_{\rm SN}$ distribution of the same SNe
(Table~\ref{RSNR25Re_deV_KSAD}).
The left panel of Fig.~\ref{SD25eff} illustrates that the main inconsistency is likely attributed to
the slower growth or decline (in case of 91bg-like events) of the SN surface density
at the central region of hosts with the radius
of about one tenth of the optical radius of galaxies (gray shaded region in the figure).
In our sample, the mean $R_{\rm e}/R_{25}$ is about four, and a similar behavior of
the $R_{\rm e}$-normalized surface density is seen at the central $0.4 R_{\rm e}$ region
(gray shaded region in the right panel of Fig.~\ref{SD25eff}).

It is important to note that different SN surveys are biased against the discovery
of SNe near the centers of host galaxies  \citep[e.g.][]{2011MNRAS.412.1419L}.
This happens because central SNe have lower contrast
with respect to the bright and often overexposed background of elliptical hosts,
increasing the difficulty of their detection in a scan of the survey figures
\citep[e.g.][]{1999AJ....117.1185H}.
In addition, host galaxy internal extinction $A_{V}<0.2$~mag exists only within the central region,
while $A_{V}$ is almost zero outside that region till to the end of optical radius of
an elliptical galaxy \citep[][]{2015A&A...581A.103G}.
Since 91bg-like events have peak luminosities that are $\sim2$ magnitudes lower than do
normal SNe~Ia \citep[e.g.][and references therein]{2008MNRAS.385...75T},
91bg-like SNe are more strongly affected by
these effects than are normal Type Ia SNe (as seen in Fig.~\ref{SD25eff}).

We now exclude SNe from the central regions of hosts
($\widetilde{R}_{\rm SN} \geq 0.1$, $\widehat{R}_{\rm SN} \geq 0.4$)
and compare the SN distributions
with the best-fitting inner-truncated de~Vaucouleurs profiles.
From the $P$-values in Table~\ref{RSNR25Re_deV_KSAD},
we see that all the inconsistencies vanish.
In Fig.~\ref{SD25eff}, we show the inner-truncated de~Vaucouleurs profiles,
extended to the central regions of host galaxies,\footnote{For illustrative purpose,
in the left panel of Fig.~\ref{SD25eff} we also present the best-fitting inner-truncated
exponential profile (black thin line, i.e. $n=1$ in Eq.~[\ref{Sersiclow}]).
The one-sample KS and AD tests show that the surface density distribution of SNe~Ia
is strongly inconsistent with the global ($P_{\rm KS}=0.050$, $P_{\rm AD}=0.005$)
and inner-truncated ($P_{\rm KS}=0.090$ [barely inconsistency], $P_{\rm AD}=0.039$) exponential models.}
and the global surface density distributions of SNe,
enabling to roughly estimate the loss in SNe~Ia discoveries, most expressive for 91bg-like events,
compared with their expected densities.
The mean loss of SNe in the central regions of elliptical galaxies is
$22\pm4$ per cent of the expected total number of Type Ia SNe.
This value is in good agreement with the similar estimation of $23\pm12$ per cent in E--S0
galaxies by \citet{1997ASIC..486...77C}, though a different method and sample were used in their study.
In our sample, the mean central losses of normal and 91bg-like SNe are $17\pm5$ and
$27\pm7$ per cent, respectively.

In addition, we check the dependence of the described bias, i.e. the central loss of SNe,
on the distances of their host galaxies \citep[the \emph{Shaw effect};][]{1979A&A....76..188S},
splitting the sample between near (${\leq {\rm 100~Mpc}}$) and far (${> {\rm 100~Mpc}}$) objects.
This separation is done to have adequate numbers of objects in the subsamples.
The surface density distributions of SNe in these distance bins show the equivalent central losses of SNe.
In this sense, it is well known that the Shaw effect is important for photographic searches and
negligible for visual/CCD searches \citep*[e.g.][]{2000ApJ...530..166H}.
Similarly, the Shaw effect is negligible in our sample, in which $\sim94$ per cent of SNe~Ia are discovered
via visual and CCD searches (see Section~\ref{samplered}).

\subsection{SNe~Ia locations vs. fractional radial light distributions of elliptical hosts}
\label{RESults3}

\begin{table}
  \centering
  \begin{minipage}{84mm}
  \caption{Consistency of the $Fr_g$ (or inner-truncated $\widecheck{Fr}_g$)
           distributions of SNe~Ia with the surface brightness distribution of elliptical host galaxies.}
  \tabcolsep 12.5pt
  \label{Frg_line_KSAD}
  \begin{tabular}{lrcc}
  \hline
    \multicolumn{1}{c}{SN~subclass ($Fr_g$ or $\widecheck{Fr}_g$)}&
    \multicolumn{1}{c}{$N_{\rm SN}$}&\multicolumn{1}{c}{$P_{\rm KS}$}&
    \multicolumn{1}{c}{$P_{\rm AD}$}\\
  \hline
    all ($Fr_g$) &94&\textbf{0.021}&0.054\\
    normal ($Fr_g$) &58&0.405&0.436\\
    91bg-like ($Fr_g$) &34&\textbf{0.044}&0.056\\
  \\
    all ($\widecheck{Fr}_g$) &79&0.482&0.404\\
    normal ($\widecheck{Fr}_g$) &46&0.758&0.719\\
    91bg-like ($\widecheck{Fr}_g$) &31&0.286&0.257\\
  \hline
  \end{tabular}
  \parbox{\hsize}{\emph{Notes.} The $P_{\rm KS}$ and $P_{\rm AD}$ are the probabilities from
                  one-sample KS and AD tests, respectively, that the distributions of $Fr_g$
                  (or inner-truncated $\widecheck{Fr}_g$) are drawn
                  from the surface brightness distribution of host galaxies.
                  The statistically significant deviations ($P$-values $\leq 0.05$) are highlighted in bold.
                  Recall that 15 SNe, which are located far outside the elliptical apertures where fluxes are
                  consistent with the sky background values, are removed from the fractional radial flux analysis
                  (see Section~\ref{samplered}).}
\end{minipage}
\end{table}

In the analysis above, we fixed the S\'{e}rsic index to $n=4$ in Eq.~[\ref{Sersiclow}]
when describing the surface density distribution of SNe~Ia,
while different elliptical host galaxies have
$n \approx 2$ to 6 in the SDSS $g$-band \citep[e.g.][]{2010ApJS..186..427N,2013MNRAS.435..623V}.
Fortunately, the distribution of fractional radial $g$-band fluxes of SNe
($Fr_g$, see Section~\ref{samplered} for definition)
allows to compare the distribution of SNe with respect to
that of the $g$-band light of elliptical host galaxies,
irrespective of their different S\'{e}rsic indices and elongations \citep[e.g.][]{2008MNRAS.388L..74F}.
If the SNe~Ia are equally likely to arise from any part of the projected light distribution of the host galaxies,
i.e. the surface brightness of galaxy $I$ and the surface density of SNe $\Sigma$ are related by
$\Sigma = {\rm Const} \times I$, then one would expect that the $Fr_g$ values are
evenly distributed throughout the projected radii of hosts (a flat distribution, independent of radius)
and the $\langle Fr_g \rangle =0.5$ \citep[e.g.][]{2006A&A...453...57J}.
For the $Fr_g$ values (from 0 to 1), the PDF and CDF are
\begin{flalign}
& p(Fr_g)=1 \, \, \, \, {\rm and} \, \, \, \, E(Fr_g)=Fr_g\, ,&
\label{pdfcdfFrg}
\end{flalign}
respectively.

\begin{figure}
\begin{center}$
\begin{array}{@{\hspace{0mm}}c@{\hspace{0mm}}}
\includegraphics[width=0.9\hsize]{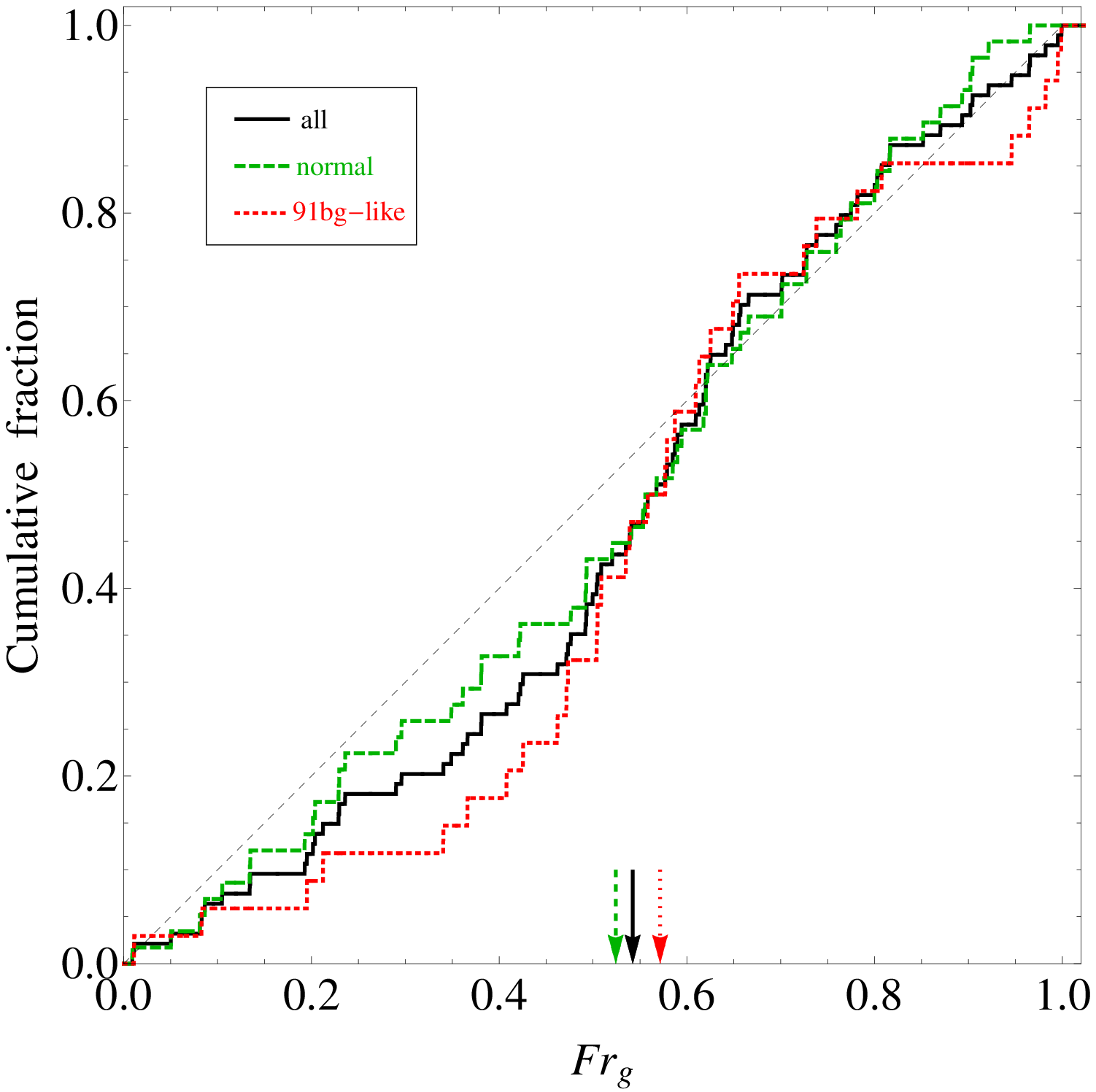}\\
\includegraphics[width=0.9\hsize]{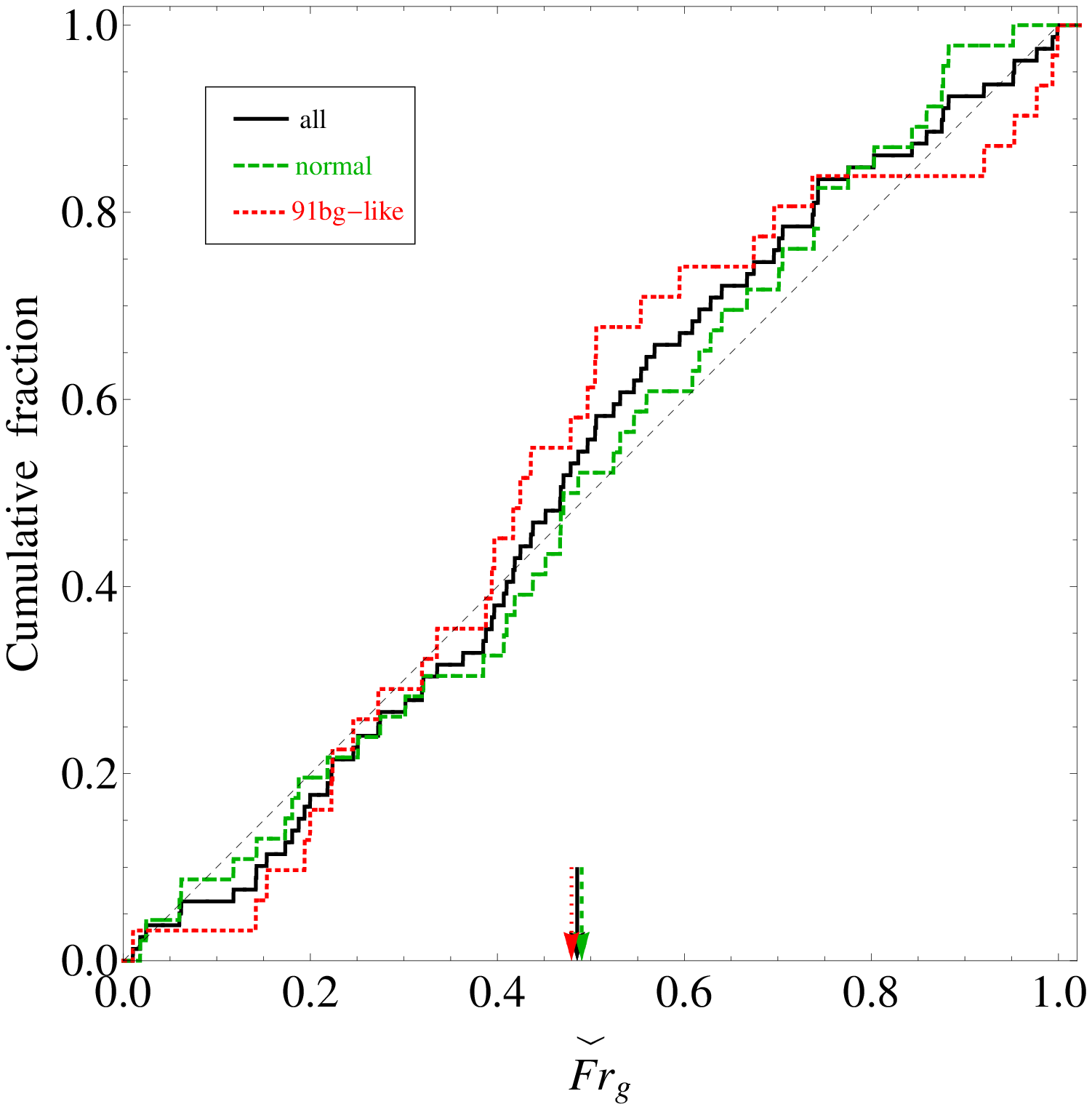}
\end{array}$
\end{center}
\caption{Upper panel: cumulative $Fr_g$ distributions of SNe~Ia
        (all -- black solid, normal -- green dashed, and 91bg-like -- red dotted)
        with respect to the $g$-band surface brightness distribution of their elliptical host galaxies
        (black thin diagonal line).
        The mean values of the distributions are shown by arrows.
        Bottom panel: same as in upper panel but for the inner-truncated $\widecheck{Fr}_g$ distributions.}
\label{Frgcumdis}
\end{figure}

From the $P$-values of one-sample KS and AD tests in Table~\ref{Frg_line_KSAD},
we see that the $Fr_g$ distribution of
Type Ia SNe is not consistent with the $g$-band light distribution of elliptical host galaxies
(for the KS statistic but marginally so in the AD statistic),
mainly due to the distribution of 91bg-like events.
The upper panel of Fig.~\ref{Frgcumdis} illustrates that, as already stated above,
the main inconsistency is due to the selection effect against the discovery of SNe~Ia
near the center of the host galaxies \citep[also seen in the right-hand panels of fig.~2 in][]{2008MNRAS.388L..74F}.
Therefore, we also use the inner-truncated fractional radial $g$-band fluxes of SNe ($\widecheck{Fr}_g$),
excluding the central region of galaxies with one tenth of the optical radius ($0.1\, R_{25}$):
\begin{flalign*}
& \widecheck{Fr}_g =
 \dfrac{Fr_g - fr_g}
 {1 - fr_g}\, ,&
\end{flalign*}
where $fr_g$ is the fractional flux of $0.1\, R_{25}$ region.
A similar definition of inner-truncated fractional flux for SNe in elliptical galaxies
was first used by \citet[][]{1976ApJ...204..519M}.

Simply replacing $Fr_g$ with $\widecheck{Fr}_g$ in Eqs.~[\ref{pdfcdfFrg}] and
using one-sample KS and AD tests, we see that
the $\widecheck{Fr}_g$ distributions of all subclasses of Type Ia SNe are
now consistent with the $g$-band light distribution of hosts,
with mean values of $\widecheck{Fr}_g$ near 0.5 as predicted
(see Table~\ref{Frg_line_KSAD} and the bottom panel of Fig.~\ref{Frgcumdis}).

\begin{table}
  \centering
  \begin{minipage}{84mm}
  \caption{Comparison of the distributions of $\widetilde{R}_{\rm SN}$, $\widehat{R}_{\rm SN}$, $Fr_g$ and $\widecheck{Fr}_g$ values
           between the subsamples of normal and 91bg-like SNe.}
  \tabcolsep 2.6pt
  \label{RSNR25Re_KSAD}
  \begin{tabular}{lrccrccc}
  \hline
    \multicolumn{1}{c}{Parameter}&\multicolumn{2}{c}{normal}&\multicolumn{1}{c}{vs.}&
    \multicolumn{2}{c}{91bg-like}&\multicolumn{1}{c}{$P_{\rm KS}$}&
    \multicolumn{1}{c}{$P_{\rm AD}$}\\
    &\multicolumn{1}{c}{$N_{\rm SN}$}&\multicolumn{1}{c}{$\langle$Parameter$\rangle$}&&
    \multicolumn{1}{c}{$N_{\rm SN}$}&\multicolumn{1}{c}{$\langle$Parameter$\rangle$}&&\\
  \hline
    $\widetilde{R}_{\rm SN}\geq0$&66&$0.43\pm0.05$& vs. &41&$0.50\pm0.07$&0.456&0.357\\
    $\widetilde{R}_{\rm SN}\geq0.1$&54&$0.52\pm0.06$& vs. &38&$0.53\pm0.07$&0.700&0.804\\
  \\
    $\widehat{R}_{\rm SN}\geq0$&66&$1.77\pm0.23$& vs. &41&$2.25\pm0.32$&0.502&0.232\\
    $\widehat{R}_{\rm SN}\geq0.4$&52&$2.18\pm0.26$& vs. &38&$2.41\pm0.33$&0.852&0.820\\
  \\
    $Fr_g$&58&$0.52\pm0.03$& vs. &34&$0.57\pm0.04$&0.606&0.429\\
    $\widecheck{Fr}_g$&46&$0.49\pm0.04$& vs. &31&$0.48\pm0.05$&0.677&0.383\\
  \hline
  \end{tabular}
  \parbox{\hsize}{\emph{Notes.} The $P_{\rm KS}$ and $P_{\rm AD}$ are the probabilities from two-sample KS and AD tests,
                  respectively, that the two distributions being compared (with respective mean values) are drawn from the same
                  parent distribution.
                  For the global distribution of all 109 SNe~Ia, the mean values of
                  $\widetilde{R}_{\rm SN}=0.45\pm0.04$ and $\widehat{R}_{\rm SN}=1.94\pm0.18$.
                  For 94 SNe~Ia, the mean value of $Fr_g=0.54\pm0.03$.
                  Recall that 15 SNe, which are located far outside the elliptical apertures where fluxes are
                  consistent with the sky background values, are removed from the fractional radial flux analysis
                  (see Section~\ref{samplered}).}
\end{minipage}
\end{table}

We now compare, in Table~\ref{RSNR25Re_KSAD}, the distributions of
$\widetilde{R}_{\rm SN}$, $\widehat{R}_{\rm SN}$, $Fr_g$ and $\widecheck{Fr}_g$ values
between the subsamples of normal and 91bg-like SNe, using the two-sample KS and AD tests.
The mean values of the distributions are also listed.
With the tests, we see no statistically significant differences between the global radial
distributions of the SN subclasses.
Similar results hold true for the inner-truncated distributions of SNe~Ia.

\subsection{The global properties of SNe~Ia elliptical host galaxies}
\label{RESults4}

\begin{table}
  \centering
  \begin{minipage}{84mm}
  \caption{Comparison of the distributions of absolute magnitudes,
           colours, sizes, elongations, stellar masses, average metallicities and luminosity-weighted ages between
           the subsamples of host galaxies of normal and 91bg-like SNe.}
  \tabcolsep 3pt
  \label{KSADforHosts}
  \begin{tabular}{lrcrcc}
  \hline
    \multicolumn{1}{c}{Parameter}&\multicolumn{1}{c}{normal}&\multicolumn{1}{c}{vs.}&\multicolumn{1}{c}{91bg-like}&
    \multicolumn{1}{c}{$P_{\rm KS}$}&\multicolumn{1}{c}{$P_{\rm AD}$}\\
    &\multicolumn{1}{c}{$\langle$Parameter$\rangle \pm \sigma$}&&
    \multicolumn{1}{c}{$\langle$Parameter$\rangle \pm \sigma$}&&\\
  \hline
    \multicolumn{6}{c}{\underline{$\widetilde{R}_{\rm SN}\geq0$ (66 vs. 41 hosts)}}\\
    $M_u$~(mag) &$-19.6\pm1.0$& vs. &$-19.8\pm0.9$&0.218&0.276\\
    $M_g$~(mag) &$-21.2\pm1.1$& vs. &$-21.5\pm0.9$&0.112&0.137\\
    $M_r$~(mag) &$-22.0\pm1.1$& vs. &$-22.3\pm0.9$&0.113&0.134\\
    $M_i$~(mag) &$-22.4\pm1.1$& vs. &$-22.7\pm0.9$&0.188&0.153\\
    $M_z$~(mag) &$-22.6\pm1.1$& vs. &$-22.9\pm0.9$&0.260&0.156\\
    $u-r$~(mag) &$2.4\pm0.1$& vs. &$2.5\pm0.1$&\textbf{0.013}&\textbf{0.007}\\
    $g-i$~(mag) &$1.2\pm0.1$& vs. &$1.2\pm0.1$&0.101&0.255\\
    $r-z$~(mag) &$0.7\pm0.05$& vs. &$0.7\pm0.04$&0.107&0.179\\
    $R_{25}$~(kpc) &$23.0\pm12.6$& vs. &$25.6\pm11.3$&0.096&0.142\\
    $R_{\rm e}$~(kpc) &$6.0\pm3.8$& vs. &$6.0\pm3.0$&0.296&0.345\\
    $a/b$ &$1.3\pm0.2$& vs. &$1.3\pm0.2$&0.766&0.729\\
    $\log(M_{\ast}/{\rm M_{\odot}})$ &$11.1^{+0.3}_{-1.3}$& vs. &$11.2^{+0.2}_{-0.6}$&0.107&0.175\\
    $\log(Z_{\ast}/{\rm Z_{\odot}})$ &$0.09^{+0.07}_{-0.08}$& vs. &$0.11^{+0.06}_{-0.07}$&0.107&0.175\\
    \emph{age}~(Gyr) &$11.7^{+2.3}_{-2.8}$& vs. &$12.8^{+1.2}_{-1.6}$&\textbf{0.017}&\textbf{0.012}\\
    \\
    \multicolumn{6}{c}{\underline{$\widetilde{R}_{\rm SN}\geq0.1$ (54 vs. 38 hosts)}}\\
    $M_u$~(mag) &$-19.6\pm1.0$& vs. &$-19.8\pm0.8$&0.286&0.416\\
    $M_g$~(mag) &$-21.2\pm1.1$& vs. &$-21.4\pm0.8$&0.309&0.275\\
    $M_r$~(mag) &$-22.0\pm1.1$& vs. &$-22.2\pm0.9$&0.306&0.252\\
    $M_i$~(mag) &$-22.4\pm1.1$& vs. &$-22.6\pm0.9$&0.365&0.325\\
    $M_z$~(mag) &$-22.7\pm1.1$& vs. &$-22.9\pm0.9$&0.365&0.314\\
    $u-r$~(mag) &$2.4\pm0.1$& vs. &$2.5\pm0.1$&\textbf{0.047}&\textbf{0.018}\\
    $g-i$~(mag) &$1.2\pm0.1$& vs. &$1.2\pm0.1$&0.480&0.721\\
    $r-z$~(mag) &$0.7\pm0.05$& vs. &$0.7\pm0.04$&0.274&0.462\\
    $R_{25}$~(kpc) &$23.3\pm12.7$& vs. &$24.2\pm9.9$&0.389&0.399\\
    $R_{\rm e}$~(kpc) &$6.1\pm4.0$& vs. &$5.5\pm2.3$&0.450&0.390\\
    $a/b$ &$1.2\pm0.2$& vs. &$1.3\pm0.2$&0.706&0.559\\
    $\log(M_{\ast}/{\rm M_{\odot}})$ &$11.2^{+0.3}_{-1.2}$& vs. &$11.2^{+0.2}_{-0.5}$&0.224&0.385\\
    $\log(Z_{\ast}/{\rm Z_{\odot}})$ &$0.09^{+0.07}_{-0.08}$& vs. &$0.10^{+0.06}_{-0.06}$&0.224&0.385\\
    \emph{age}~(Gyr) &$11.9^{+2.1}_{-2.7}$& vs. &$12.7^{+1.3}_{-1.7}$&\textbf{0.025}&\textbf{0.032}\\
  \hline
  \end{tabular}
  \parbox{\hsize}{\emph{Notes.} The $P_{\rm KS}$ and $P_{\rm AD}$ are the probabilities from two-sample KS and AD tests,
                  respectively, that the two distributions being compared (with respective mean values and standard deviations)
                  are drawn from the same parent distribution.
                  The statistically significant differences ($P$-values $\leq 0.05$) between the distributions are highlighted in bold.}
\end{minipage}
\end{table}

In the SDSS DR15, different estimates of the parameters of galaxies
(e.g. stellar mass, metallicity and age of stellar population)
encompass calculations based on various stellar population models
(e.g. Evolutionary Population Synthesis, \citealt{2005MNRAS.362..799M};
Principal Component Analysis-based model, \citealt{2012MNRAS.421..314C};
Flexible Stellar Population Synthesis, \citealt*{2009ApJ...699..486C}),
and different assumptions about galaxy extinction and star formation
histories.\footnote{For more detailed information with corresponding references, the reader is referred to
\href{https://www.sdss.org/dr15/spectro/galaxy/}{https://www.sdss.org/dr15/spectro/galaxy/}.}
However, from 109 SNe~Ia elliptical hosts of our study,
only 43 SNe (29 normal, thirteen 91bg-like and one 06gz-like) have
available SDSS spectra of hosts, thus reliable estimates of mass, age, and metallicity.
Therefore, instead of using them we prefer to estimate the stellar masses ($M_{\ast}$) of all our elliptical hosts,
using the empirical relation of \citet[][]{2011MNRAS.418.1587T} between $\log(M_{\ast}/{\rm M_{\odot}})$,
$g-i$ colour and $i$-band absolute magnitude ($M_i$)
as determined from more than $10^5$ galaxies with redshifts $z<0.65$:
\begin{flalign}
& \log\Bigl(\dfrac{M_{\ast}}
 {\rm M_{\odot}}\Bigr)=1.15+0.70(g-i)-0.4M_i\, ,&
\end{flalign}
where $M_{\ast}$ has solar mass units.
According to \citeauthor{2011MNRAS.418.1587T},
this relation provides an estimate of the stellar mass-to-light ratio
($M_{\ast}/L_i$) to a $1\sigma$ accuracy of $\sim0.1$~dex.
In addition, to estimate average host galaxy stellar metallicities,
we use the \citet[][]{2006MNRAS.370.1106G} correlation between stellar mass
of E--S0 galaxy and $\log(Z_{\ast}/{\rm Z_{\odot}})$
as determined from about 26000 SDSS galaxies
\citep[see also][for early-type/high mass galaxies]{2017MNRAS.472.2833S}:
\begin{flalign}
& \log\Bigl(\dfrac{Z_{\ast}}
 {\rm Z_{\odot}}\Bigr)=-1.757+0.168\log\Bigl(\dfrac{M_{\ast}}{\rm M_{\odot}}\Bigr)\, ,&
\end{flalign}
where $Z_{\ast}$ has solar metallicity units, with a scatter of $\sim0.1$~dex.
It should be noted that we use mass measurements coupled to a (monotonic) formula
to convert it to host metallicity, which adds no original information to
the statistical analysis, however, this gives a chance to qualitatively discuss our
results in term of metallicities of SNe~Ia hosts (see Section~\ref{DISconc}).

\begin{figure}
\begin{center}$
\begin{array}{@{\hspace{0mm}}c@{\hspace{0mm}}}
\includegraphics[width=1\hsize]{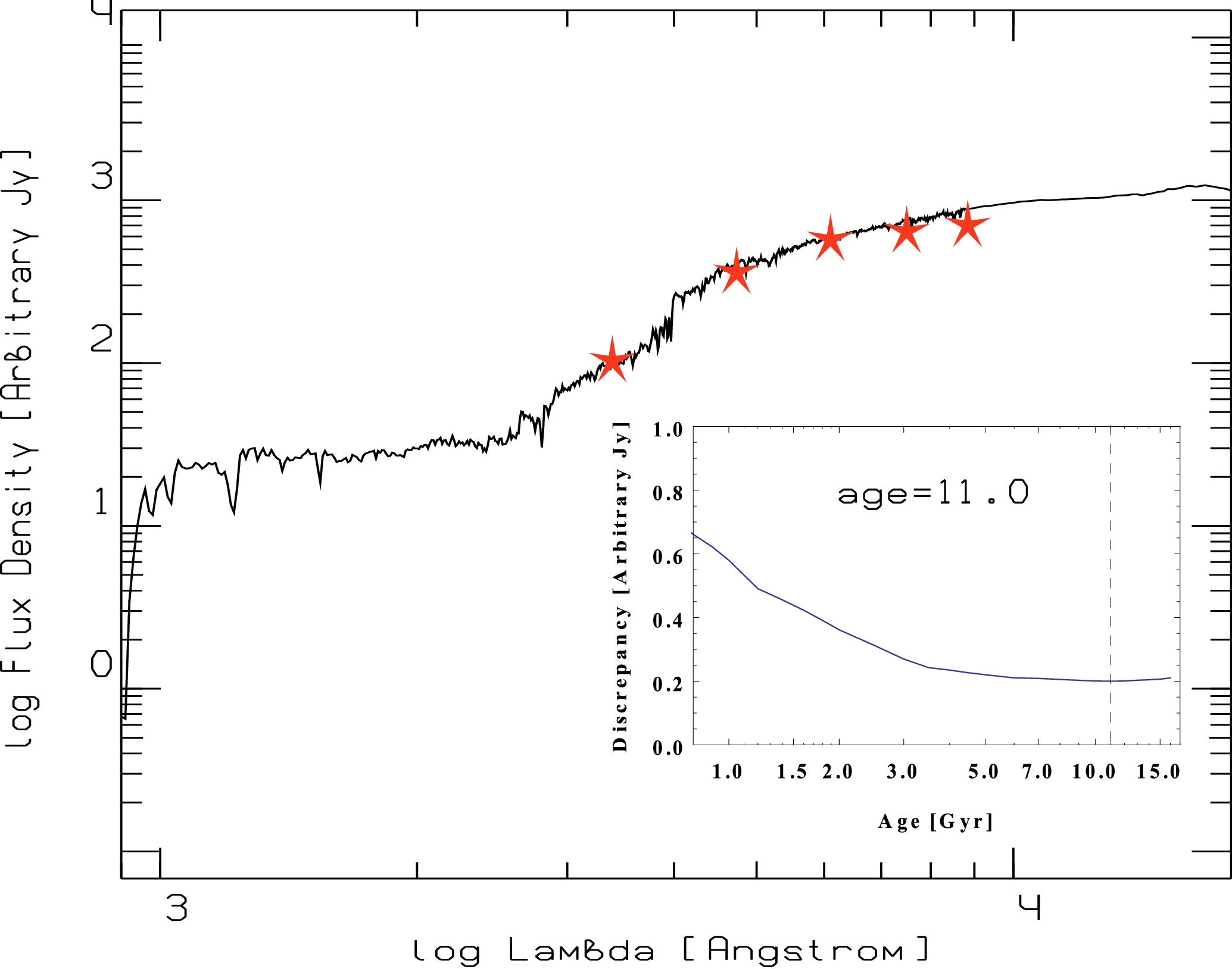}
\end{array}$
\end{center}
\caption{Photometric points (in the SDSS five bands, red asterisks)
         of SN~2018zs host elliptical galaxy with the best-SED, all in the rest-frame.
         The inset shows the curve of the dependence of the model rms deviations
         on age for the galaxy, with the best-age of 11~Gyr
         (minimum of the curve is shown by the vertical dashed line).}
\label{hgalSN2018zs}
\end{figure}

Finally, following the procedure outlined in \citet[][]{2000A&AT...19..662V},
we use the fixed redshifts of SN hosts to fit the PEGASE.2 \citep[][]{1997A&A...326..950F,1999astro.ph.12179F}
elliptical galaxy models to our $u$-, $g$-, $r$-, $i$- and $z$-band photometry
to determine the luminosity-weighted ages of hosts.\footnote{The luminosity-weighted ages of our
104 elliptical host galaxies are available in the online version (Supporting Information) of this article.}
In short, the measured five photometric points of a host galaxy
in the SDSS bands with fixed redshift are used to select the best location
of the points on the spectral energy distribution (SED) templates.
Such a location can be found by shifting
the points lengthwise and transverse the SED template at which the sum of the squares of
the discrepancies is a minimum. From the PEGASE.2 model
\citep[][]{1997A&A...326..950F,1999astro.ph.12179F}, the procedure uses already
computed collection of synthetic SED templates for different ages
(up to 19~Gyr) of elliptical galaxies.
Fig.~\ref{hgalSN2018zs} presents an example of SN host galaxy photometric points
(in the SDSS five bands) with the best-SED, all in the rest-frame.
For more detailed information on the SED
fitting procedure with filter smoothing option,
the reader is referred to \href{http://sed.sao.ru/}{http://sed.sao.ru/}.

\begin{figure*}
\begin{center}$
\begin{array}{@{\hspace{0mm}}c@{\hspace{0mm}}}
\includegraphics[width=0.75\hsize]{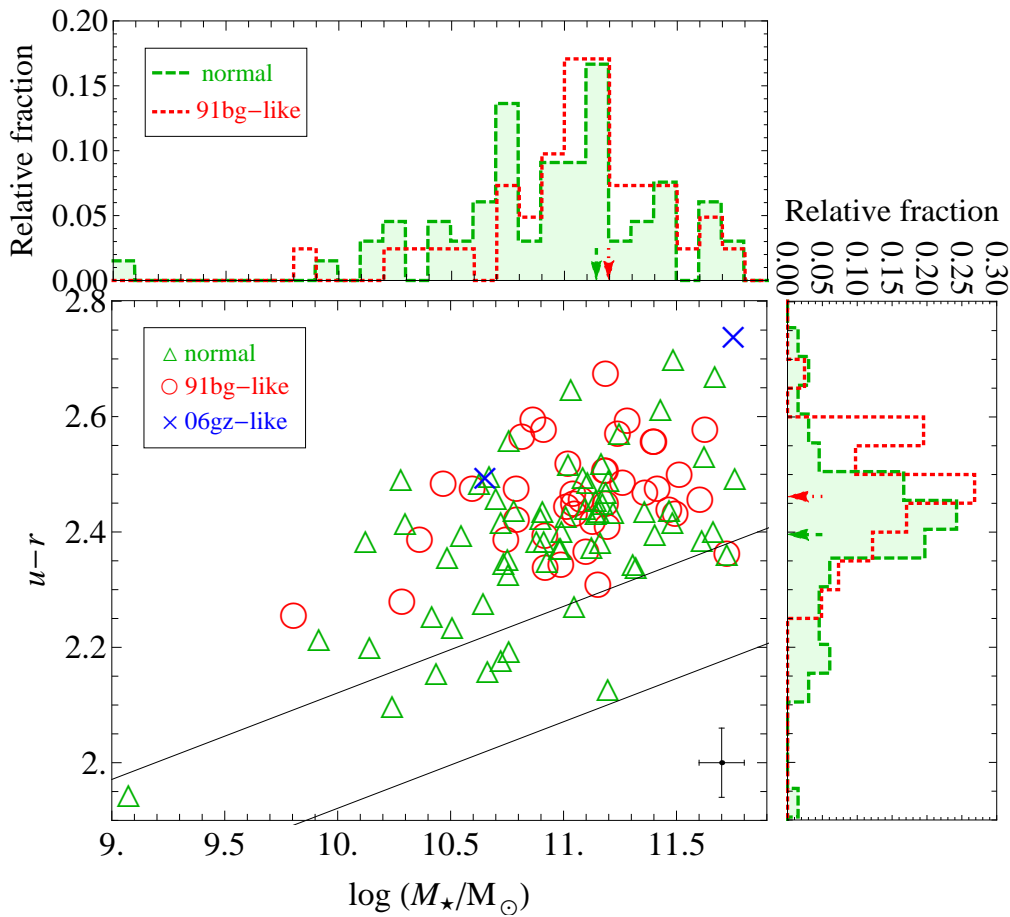}
\end{array}$
\end{center}
\caption{The $u-r$ colour--mass diagram for 109 SNe~Ia elliptical host galaxies.
         Green triangles, red circles and blue crosses show normal, 91bg-like and 06gz-like SNe hosts, respectively.
         The region between two solid lines indicates the Green Valley
         (see the text for more details).
         The vertical and horizontal error bars, in the bottom-right corner, show the characteristic errors
         in the colour and mass estimations, respectively.
         For normal (green dashed and filled) and 91bg-like (red dotted) SNe hosts,
         the right and upper panels represent separately the histograms of the colours and masses, respectively.
         The mean values of the distributions are shown by arrows.}
\label{colMag3}
\end{figure*}

\begin{figure}
\begin{center}$
\begin{array}{@{\hspace{0mm}}c@{\hspace{0mm}}}
\includegraphics[width=1\hsize]{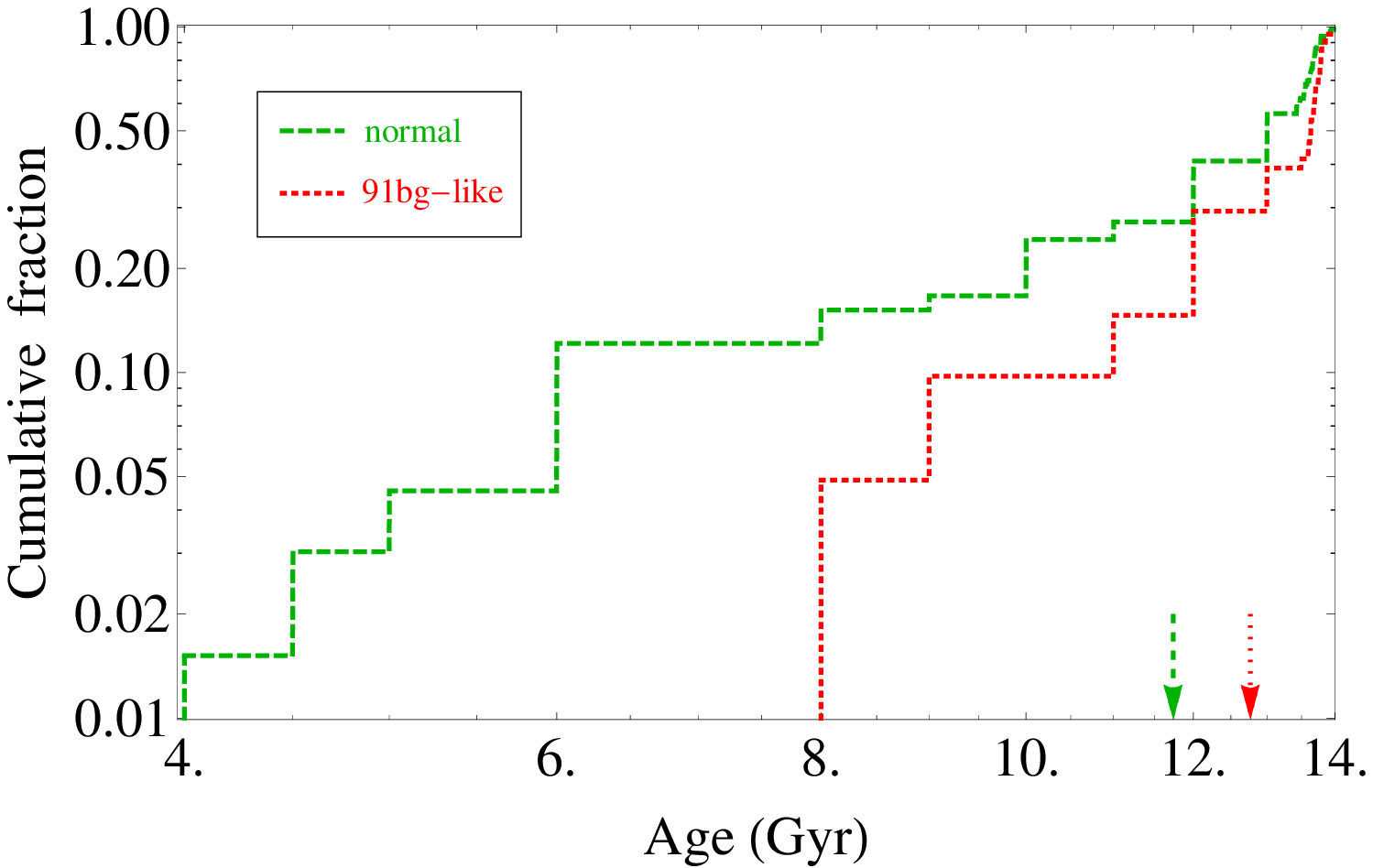}
\end{array}$
\end{center}
\caption{Cumulative distributions of luminosity-weighted ages of elliptical host
         galaxies of normal (green dashed) and 91bg-like (red dotted) SNe.
         The mean values of the distributions are shown by arrows.}
\label{Agenorm91bg}
\end{figure}

To reveal possible differences in global properties of SNe~Ia elliptical hosts,
in Table~\ref{KSADforHosts}, using the two-sample KS and AD tests, we compare absolute magnitudes,
colours, sizes, elongations, stellar masses, average metallicities and luminosity-weighted ages between
the subsamples of host galaxies of normal and 91bg-like SNe.
The table shows that the distributions of absolute magnitudes, $g-i$ and $r-z$ colours (red part of the SEDs),
sizes, elongations, stellar masses and average metallicities are not significantly different between host galaxies of
normal and 91bg-like SNe.
On the other hand, the distributions of $u-r$ colours (blue part of the SEDs) and
luminosity-weighted ages of the hosts are significantly inconsistent between the subclasses of SNe~Ia.
In the histograms of Fig.~\ref{colMag3}, we show the distributions of host galaxy stellar masses and $u-r$ colours.
The cumulative distributions of luminosity-weighted ages of the elliptical hosts are presented in Fig.~\ref{Agenorm91bg}.
It is clear that, despite their comparable stellar masses, the elliptical host galaxies of normal SNe~Ia are
on average bluer and younger than those of 91bg-like SNe.

In Table~\ref{KSADforHosts}, we also check the impact of the described bias in Subsection~\ref{RESults2},
i.e. the stronger central loss of 91bg-like SNe, on the comparison of the global properties of
ellipticals by excluding the host galaxies with $\widetilde{R}_{\rm SN}<0.1$.
We obtain nearly identical results showing that the central bias has negligible impact on the comparison of
the elliptical host galaxies in Table~\ref{KSADforHosts}.

\section{Discussion and summary}
\label{DISconc}

In this section, we discuss all the results obtained above and give summary within
an evolutionary (interacting) scenario of SNe~Ia elliptical host galaxies that
can explain the similarity of the spatial distributions of normal and 91bg-like SNe
in hosts and at the same time the differences of some global properties of elliptical hosts
such as the $u-r$ colours and the ages of the stellar population.

In Subsection~\ref{RESults2}, we have shown that the distributions of projected galactocentric radii
(with different normalizations) of normal and 91bg-like SNe in elliptical galaxies follow
the de~Vaucouleurs model,
except in the central region of ellipticals where the different SN surveys are biased
against the discovery of the events (Table~\ref{RSNR25Re_deV_KSAD} and Fig.~\ref{SD25eff}).
These results are in agreement with a more generalized result of \citet{2008MNRAS.388L..74F},
who showed that the projected surface density distribution of Type Ia SNe
(without separating the subclasses) in morphologically selected early-type host galaxies
is consistent with the de~Vaucouleurs profile \citep[see also][]{2004AstL...30..729T,2010ApJ...715.1021D}.
Even without specifying the profile shape and excluding the bias against central SNe,
the radial distributions of SN~Ia subclasses are consistent with the radial light distribution of
stellar populations of elliptical hosts in the SDSS $g$-band
(Table~\ref{Frg_line_KSAD} and Fig.~\ref{Frgcumdis}).
We have not seen any significant differences between the radial distributions
of normal and 91bg-like SNe (Table~\ref{RSNR25Re_KSAD}).

These results are in agreement with those of \citet[][]{2005ApJ...634..210G},
who studied the distribution of 57 local Type Ia SNe LC decline rates
($\Delta m_{15}$) in the $B$-band versus projected distances (in kpc)
from the centers of spiral and E--S0 host galaxies.
Despite their smaller statistics of E--S0 galaxies, they found that the $\Delta m_{15}$ values
are distributed evenly with projected galactocentric radii, showing no preference to the center of
host galaxies for slowly declining (normal SNe~Ia) or faster declining (91bg-like) SNe
\citep[see also][for projected and normalized galactocentric radii]{2000ApJ...542..588I}.
Using a larger SN~Ia sample at redshifts below 0.25 and output parameters from two LC fitters,
MLCS2k2 \citep*{2007ApJ...659..122J} and SALT2 \citep{2007A&A...466...11G}, \citet{2012ApJ...755..125G}
also studied the dependencies between SN properties and the projected galactocentric radii.
For 64 SNe~Ia in elliptical hosts, with determined morphology based on the concentration indices and S\'{e}rsic profiles,
the authors found some indications that SNe tend to have faster declining LCs if they explode
at larger galactocentric radii. However, this trend is visible when the LC parameters from MLCS2k2 were used,
in contrast to the homologous parameters from SALT2.
In addition, \citeauthor{2012ApJ...755..125G} noted that
their finding might be due to the possible selection effects and
explained by the difficulty in detecting faster declining/fainter SNe~Ia near the galaxy center,
which we demonstrated in Subsection~\ref{RESults2}, based on the surface density distributions of
normal and 91bg-like events in elliptical hosts.

In Subsection~\ref{RESults4}, we have shown that the distributions of absolute magnitudes,
stellar masses and average metallicities are not significantly different between host galaxies of normal
and 91bg-like SNe (Table~\ref{KSADforHosts}).
Similar results were also obtained by \citet[][]{2005ApJ...634..210G}, who found
no correlation between the LC decline rates of SNe~Ia
and absolute $B$-band magnitudes (a sufficient tracer of galactic mass) of their E--S0 hosts.
\citet[][]{2008ApJ...685..752G} also studied optical absorption-line spectra of 29 early-type (mostly E--S0)
host galaxies of SNe~Ia up to about 200~Mpc and found a mild correlation, if any,
between host global metallicity and SN~Ia peak luminosity.

Indeed, the variety of metallicities of the main-sequence stars that become white dwarfs
could theoretically affect the mass of $^{56}{\rm Ni}$ synthesized in SNe~Ia \citep*{2003ApJ...590L..83T},
and cause a variety in the properties of SNe~Ia (e.g. in luminosities and/or decline rates).
These authors hypothesized that less luminous SNe~Ia arise from high-metallicity progenitors
that produce less $^{56}{\rm Ni}$. However, \citet{2009ApJ...691..661H} noted that the effect is dominant at
metallicities significantly above solar, whereas early-type hosts of SNe~Ia
have only moderately above-solar metallicities (with no detectable star formation).
In this respect, our elliptical host galaxies also span moderately above-solar metallicities
(see Table~\ref{KSADforHosts}),
mostly within $0 \lesssim \log(Z_{\ast}/{\rm Z_{\odot}}) \lesssim 0.2$ range,
and therefore the metallicity effect in our sample might be sufficient to vary the optical peak brightness of
SN~Ia by less than 0.2~mag \citep{2003ApJ...590L..83T}, but not enough for the differences between peak magnitudes
of normal and 91bg-like SNe \citep[as already mentioned, the latter have peak luminosities that
are $2\pm0.5$~mag lower in optical bands than do normal SNe, see][]{2008MNRAS.385...75T}.

On the other hand, the radial metallicity gradient in elliptical galaxies \citep[e.g.][]{1999PASP..111..919H}
might be a useful tool to probe the differences between the properties of SN~Ia subclasses.
However, \citet{2015A&A...581A.103G} recently studied nearby galaxies, including 41 ellipticals,
with redshifts $<0.03$, using the precise data of integral field spectroscopy, and found that
the average radial metallicity profile of elliptical galaxies (with negative gradient)
declines only moderately from 0.2~dex above solar to solar from the
galactic center up to $3R_{\rm e}$, respectively.
Therefore, most probably this small metallicity variation does not allow
\citep[according to][]{2003ApJ...590L..83T} to see the differences between
the distributions of normal and 91bg-like SNe along the radius of their elliptical hosts (Table~\ref{RSNR25Re_KSAD}).
Our results confirm that the masses as well as global and radial metallicity distributions of elliptical hosts
are not decisive factors of the nature of normal and 91bg-like SN populations
\citep[see also discussions by][]{2000ApJ...542..588I,2005ApJ...634..210G,2008ApJ...685..752G}.

At the same time, in Subsection~\ref{RESults4}, we have shown that the distributions of
$u-r$ colours and luminosity-weighted ages are inconsistent significantly between
the elliptical host galaxies of different SN~Ia subclasses (Table~\ref{KSADforHosts}):
the hosts of normal SNe~Ia are on average bluer (the right histograms in Fig.~\ref{colMag3})
and younger (Fig.~\ref{Agenorm91bg}) than those of 91bg-like SNe.
These results are in excellent agreement with those of \citet[][]{2008ApJ...685..752G}, who found
a strong correlation between SN peak luminosities and the luminosity-weighted ages
of dominant population of E--S0 hosts.
They suggested that SNe~Ia in galaxies with a characteristic age greater than several Gyr are on average
$\sim1$~mag fainter at the peak in $V$-band than those in early-type hosts with younger populations
(i.e. a fairly large number of subluminous/91bg-like SNe are discovered in older hosts).
In addition, \citet[][]{2008ApJ...685..752G} noted about the difficulty to distinguish whether this effect
is a smooth transition with age or the result of two distinct SN~Ia populations.
Most recently, \citet[][]{2019arXiv190410139P} analysed integral field observations of the apparent/underlying
explosion sites of eleven spectroscopically identified 91bg-like SNe (redshifts $\leq 0.04$) in hosts with
different morphologies (including six E--S0 galaxies)
and found that the majority of the stellar populations that host these events are dominated by old stars
with a lack of evidence for recent star formation.
\citeauthor{2019arXiv190410139P} concluded that the 91bg-like SN progenitors are likely to have
delay time distribution weighted toward long delay times \citep[$>6$~Gyr, see also][]{2017NatAs...1E.135C},
much longer than the typical delay times of normal SNe~Ia in star forming environments,
whose delay times peak between several hundred Myr and $\sim 1$~Gyr
\citep[e.g.][]{2014MNRAS.445.1898C,2014ARA&A..52..107M}.
These results are in good agreement with our findings in Table~\ref{KSADforHosts} and Fig.~\ref{Agenorm91bg}.

It is important to note that the global ages of elliptical galaxies are not significantly different,
on average, from local ones at any radii, i.e. there is no clear age gradient in ellipticals,
being only mildly negative up to $R_{\rm e}$ and flat beyond that radius \citep[e.g.][]{2015A&A...581A.103G}.
For this reason, we see no difference between the radial distributions of the subclasses of
SNe~Ia (Table~\ref{RSNR25Re_KSAD}), meanwhile seeing the clear differences of the global ages of
normal and 91bg-like hosts (Table~\ref{KSADforHosts} and Fig.~\ref{Agenorm91bg}).
Thus, our results support the earlier suggestions
\citep[e.g.][]{2000ApJ...542..588I,2005ApJ...634..210G,2008ApJ...685..752G,2016ApJS..223....7K}
that the age of SN~Ia progenitor populations is a more important factor than metallicity or
mass of elliptical host galaxies in determining the properties of normal and 91bg-like events.

We now interpret and summarise our results within an evolutionary (interacting)
scenario of SNe~Ia elliptical host galaxies.
In Fig.~\ref{colMag3}, we show $u-r$ colour--mass diagram of elliptical host galaxies
(the right and upper panels represent separately the histograms of the colours and masses, respectively).
In Fig.~\ref{colMag3}, the region between two solid lines indicates the location of the Green Valley,
i.e. the region between blue star-forming galaxies and the Red Sequence of quiescent E--S0 galaxies
\citep[e.g.][]{2011ApJ...736..110M,2014MNRAS.440..889S}.
For galaxies with elliptical morphology, this is a transitional state through which blue galaxies
evolve into the Red Sequence via major merging processes
with morphological transformation from disc to spheroidal shape
\citep[e.g.][]{2010ApJ...714L.108S,2014MNRAS.442..533M},
and/or a state of galaxies demonstrating some residual star formation
via minor merging processes with no global changes in spheroidal structure
\citep[e.g.][]{2009MNRAS.394.1713K}.

It should be noted that we use the modification of the Green Valley defined by \citet{2014MNRAS.440..889S}.
These authors used the SDSS \texttt{modelMag} values\footnote{The \texttt{modelMag} values are first calculated
using the best-fit parameters in the $r$-band, and then applied these parameters to all other SDSS bands,
therefore, the light is measured consistently through the same aperture in all bands.}
of about 9000 early- and about 17000 late-type galaxies
with redshifts $0.02 < z < 0.05$, while we use the magnitudes of host galaxies based on
the $g$-band $25~{\rm mag~arcsec^{-2}}$ elliptical apertures (see Section~\ref{samplered}).
The comparison of SDSS DR15 \texttt{modelMag} measurements of our elliptical host galaxies with those obtained in
Section~\ref{samplered}, and the best-fit of our $u-r$ versus $\log(M_{\ast}/{\rm M_{\odot}})$
bring a negative shift and a small change in slope for
the modification of the Green Valley\footnote{The best-fit is $u-r = 0.721 + 0.155\, \log(M_{\ast}/{\rm M_{\odot}})$
for normal and 91bg-like SNe hosts.
The upper and bottom borders of our Green Valley in Fig.~\ref{colMag3} are simply negative shifts of
the best-fit in 0.1 and 0.3~mag, respectively.}
in comparison with that in \citet{2014MNRAS.440..889S}.

\begin{figure}
\begin{center}$
\begin{array}{@{\hspace{0mm}}c@{\hspace{0mm}}}
\includegraphics[width=1\hsize]{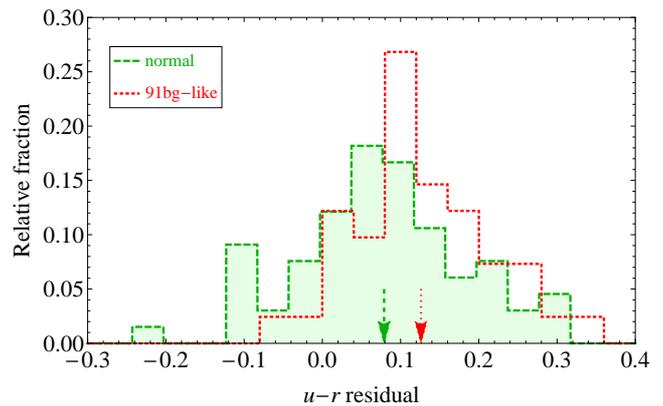}
\end{array}$
\end{center}
\caption{Distributions of colour residuals of elliptical hosts
         of normal (green dashed and filled) and 91bg-like (red dotted) SNe
         relative to the upper border of our Green Valley.
         The mean values of the distributions are shown by arrows.}
\label{urresidual}
\end{figure}

In Fig.~\ref{colMag3}, we see that the tail of the colour distribution of normal SN hosts
stretches well into the Green Valley, while the same tail of 91bg-like SN hosts barely reaches
the Green Valley border, and only at high stellar masses.
To quantify this difference, we compare the distributions of colour residuals of elliptical hosts
of the SN subclasses relative to the upper border of our Green Valley (see Fig.~\ref{urresidual}).
The two-sample KS and AD tests show that the distributions are significantly different
($P_{\rm KS}=0.049$, $P_{\rm AD}=0.026$).
Therefore, the bluer and younger elliptical hosts of normal SNe~Ia should have
more residual star formation \citep[e.g.][]{2009MNRAS.394.1713K,2014MNRAS.440..889S}
that gives rise to younger SN~Ia progenitors, resulting in normal SNe with shorter delay times
(e.g. \citealt{2014MNRAS.445.1898C,2014ARA&A..52..107M}; \citealt*{2017ApJ...850..135U}).
Interestingly, the results of \citet[][]{2016A&A...585A..92G,2016A&A...588A..68G}
reveal that in such galaxies the residual star formation is well mixed radially
and distributed within entire stellar population.

As was recalled in the Introduction, the rate of SNe~Ia
can be represented as a linear combination of prompt and delayed components
\citep[e.g.][]{2005ApJ...629L..85S}.
The prompt component is dependent on the rate of recent star formation,
and the delayed component is dependent on the galaxy total stellar mass
\citep[e.g.][]{2005A&A...433..807M,2011MNRAS.412.1473L,2011Ap.....54..301H}.
In this context, the normal SNe~Ia with shorter delay times correspond to
the prompt component.
The bluer and younger ellipticals (with residual star formation) can also produce 91bg-like events
with lower rate \citep[e.g.][]{2008ApJ...685..752G}, because of long delay times of these SNe
\citep[e.g.][]{2019arXiv190410139P}, i.e.
a delayed component of SN~Ia explosions \citep[e.g.][]{2005ApJ...629L..85S,2011ApJ...727..107G}.
However, the distribution of host ages (lower age limit of the delay times)
of 91bg-like SNe does not extend down
to the stellar ages that produce a significant excess of $u-r$ colour
(i.e. $u$-band flux, see Figs.~\ref{colMag3} and \ref{Agenorm91bg})
-- younger stars in elliptical hosts do not produce 91bg-like SNe,
i.e. the 91bg-like events have no prompt component.
The redder and older elliptical hosts that already exhausted nearly all star formation budget
during the evolution \citep[e.g.][]{2014MNRAS.440..889S} may produce significantly less normal SNe~Ia
with shorter delay times, outnumbered by 91bg-like SNe with long delay times.

Finally, we would like to note that our results favor SN~Ia progenitor models such as
helium-ignited violent mergers as a unified model for
normal (CO~WD primary with CO~WD companion)
and 91bg-like (CO~WD primary with He~WD companion)
SNe \citep[e.g.][]{2013ApJ...770L...8P,2017NatAs...1E.135C}
that have the potential to explain the different luminosities, delay times,
and relative rates of the SN subclasses (see also
\citealt*{2010Natur.463..924G,2011NewA...16..250L},
for discussions of binary WDs mergers in elliptical galaxies).
In particular, the models predict shorter delay times for normal SNe~Ia
in agreement with our finding that normal SNe occur in younger stellar population
of elliptical hosts. Moreover, the model prediction of very long delay times for
91bg-like SNe \citep[$\gtrsim$~several Gyr,][]{2017NatAs...1E.135C}
is in good qualitative agreement with our estimation of older ages of
host galaxies of these events.

In the years of the surveys by robotic telescopes on different sites on the globe
\citep[e.g. All Sky Automated Survey for SuperNovae][]{2017PASP..129j4502K}
and of the forthcoming Large Synoptic Survey Telescope \citep[][]{2002SPIE.4836...10T},
thousands of relatively nearby SNe~Ia with
spectroscopic confirmations are expected to be discovered that will provide larger
and better-defined samples of these transient events.
We will then be able to place tighter constraints on the evolutionary scenarios
of host galaxies and on the photometric and spectroscopic properties of
Type Ia SNe with different progenitor models.

\section*{Acknowledgements}

We would like to thank the referee, Michael Childress,
for excellent comments that improved the clarity of this paper.
LVB, AAH, and AGK acknowledge the hospitality of the
Institut d'Astrophysique de Paris (France) during their
stay as visiting scientists supported by
the Programme Visiteurs Ext\'{e}rieurs (PVE).
This work was supported by the RA MES State Committee of Science,
in the frames of the research project number 15T--1C129.
This work was made possible in part by a research grant from the
Armenian National Science and Education Fund (ANSEF)
based in New York, USA.
V.A. acknowledges the support from Funda\c{c}\~ao para
a Ci\^encia e Tecnologia (FCT) through Investigador FCT contract nr.
IF/00650/2015/CP1273/CT0001 and the support from FCT/MCTES through national
funds (PIDDAC) - UID/FIS/04434/2019.
Funding for the SDSS-IV has been provided by the
Alfred P.~Sloan Foundation, the US Department of Energy Office of Science,
and the Participating Institutions. SDSS-IV acknowledges
support and resources from the Center for High-Performance Computing at
the University of Utah. The SDSS web site is \href{http://www.sdss.org/}{www.sdss.org}.
SDSS-IV is managed by the Astrophysical Research Consortium for the
Participating Institutions of the SDSS Collaboration including the
Brazilian Participation Group, the Carnegie Institution for Science,
Carnegie Mellon University, the Chilean Participation Group, the French Participation Group,
Harvard-Smithsonian Center for Astrophysics,
Instituto de Astrof\'isica de Canarias, The Johns Hopkins University,
Kavli Institute for the Physics and Mathematics of the Universe (IPMU) /
University of Tokyo, the Korean Participation Group, Lawrence Berkeley National Laboratory,
Leibniz Institut f\"ur Astrophysik Potsdam (AIP),
Max-Planck-Institut f\"ur Astronomie (MPIA Heidelberg),
Max-Planck-Institut f\"ur Astrophysik (MPA Garching),
Max-Planck-Institut f\"ur Extraterrestrische Physik (MPE),
National Astronomical Observatories of China, New Mexico State University,
New York University, University of Notre Dame,
Observat\'ario Nacional / MCTI, The Ohio State University,
Pennsylvania State University, Shanghai Astronomical Observatory,
United Kingdom Participation Group,
Universidad Nacional Aut\'onoma de M\'exico, University of Arizona,
University of Colorado Boulder, University of Oxford, University of Portsmouth,
University of Utah, University of Virginia, University of Washington, University of Wisconsin,
Vanderbilt University, and Yale University.

\bibliography{SphSNIabib}

\newcommand{\SortNoop}[1]{}
\begin{thebibliography}{}
\makeatletter
\relax
\def\mn@urlcharsother{\let\do\@makeother \do\$\do\&\do\#\do\^\do\_\do\%\do\~}
\def\mn@doi{\begingroup\mn@urlcharsother \@ifnextchar [ {\mn@doi@}
  {\mn@doi@[]}}
\def\mn@doi@[#1]#2{\def\@tempa{#1}\ifx\@tempa\@empty \href
  {http://dx.doi.org/#2} {doi:#2}\else \href {http://dx.doi.org/#2} {#1}\fi
  \endgroup}
\def\mn@eprint#1#2{\mn@eprint@#1:#2::\@nil}
\def\mn@eprint@arXiv#1{\href {http://arxiv.org/abs/#1} {{\tt arXiv:#1}}}
\def\mn@eprint@dblp#1{\href {http://dblp.uni-trier.de/rec/bibtex/#1.xml}
  {dblp:#1}}
\def\mn@eprint@#1:#2:#3:#4\@nil{\def\@tempa {#1}\def\@tempb {#2}\def\@tempc
  {#3}\ifx \@tempc \@empty \let \@tempc \@tempb \let \@tempb \@tempa \fi \ifx
  \@tempb \@empty \def\@tempb {arXiv}\fi \@ifundefined
  {mn@eprint@\@tempb}{\@tempb:\@tempc}{\expandafter \expandafter \csname
  mn@eprint@\@tempb\endcsname \expandafter{\@tempc}}}

\bibitem[\protect\citeauthoryear{{Aguado} et~al.,}{{Aguado}
  et~al.}{2019}]{2019ApJS..240...23A}
{Aguado} D.~S.,  et~al., 2019, \mn@doi [\apjs] {10.3847/1538-4365/aaf651},
  \href {http://adsabs.harvard.edu/abs/2019ApJS..240...23A} {240, 23}

\bibitem[\protect\citeauthoryear{{Anderson} \& {James}}{{Anderson} \&
  {James}}{2009}]{2009MNRAS.399..559A}
{Anderson} J.~P.,  {James} P.~A.,  2009, \mn@doi [\mnras]
  {10.1111/j.1365-2966.2009.15324.x}, \href
  {http://adsabs.harvard.edu/abs/2009MNRAS.399..559A} {399, 559}

\bibitem[\protect\citeauthoryear{{Anderson}, {James}, {F{\"o}rster},
  {Gonz{\'a}lez-Gait{\'a}n}, {Habergham}, {Hamuy}  \& {Lyman}}{{Anderson}
  et~al.}{2015}]{2015MNRAS.448..732A}
{Anderson} J.~P.,  {James} P.~A.,  {F{\"o}rster} F.,  {Gonz{\'a}lez-Gait{\'a}n}
  S.,  {Habergham} S.~M.,  {Hamuy} M.,   {Lyman} J.~D.,  2015, \mn@doi [\mnras]
  {10.1093/mnras/stu2712}, \href
  {http://adsabs.harvard.edu/abs/2015MNRAS.448..732A} {448, 732}

\bibitem[\protect\citeauthoryear{{Aramyan} et~al.,}{{Aramyan}
  et~al.}{2016}]{2016MNRAS.459.3130A}
{Aramyan} L.~S.,  et~al., 2016, \mn@doi [\mnras] {10.1093/mnras/stw873}, \href
  {http://adsabs.harvard.edu/abs/2016MNRAS.459.3130A} {459, 3130}

\bibitem[\protect\citeauthoryear{{Arnett}}{{Arnett}}{1982}]{1982ApJ...253..785A}
{Arnett} W.~D.,  1982, \mn@doi [\apj] {10.1086/159681}, \href
  {https://ui.adsabs.harvard.edu/abs/1982ApJ...253..785A} {253, 785}

\bibitem[\protect\citeauthoryear{{Barbon}, {Buond{\'{\i}}}, {Cappellaro}  \&
  {Turatto}}{{Barbon} et~al.}{1999}]{1999A&AS..139..531B}
{Barbon} R.,  {Buond{\'{\i}}} V.,  {Cappellaro} E.,   {Turatto} M.,  1999,
  \mn@doi [\aaps] {10.1051/aas:1999404}, \href
  {http://adsabs.harvard.edu/abs/1999A%26AS..139..531B} {139, 531}

\bibitem[\protect\citeauthoryear{{Bartunov}, {Makarova}  \&
  {Tsvetkov}}{{Bartunov} et~al.}{1992}]{1992A&A...264..428B}
{Bartunov} O.~S.,  {Makarova} I.~N.,   {Tsvetkov} D.~I.,  1992, \aap, \href
  {http://adsabs.harvard.edu/abs/1992A%26A...264..428B} {264, 428}

\bibitem[\protect\citeauthoryear{{Benetti} et~al.,}{{Benetti}
  et~al.}{2005}]{2005ApJ...623.1011B}
{Benetti} S.,  et~al., 2005, \mn@doi [\apj] {10.1086/428608}, \href
  {http://adsabs.harvard.edu/abs/2005ApJ...623.1011B} {623, 1011}

\bibitem[\protect\citeauthoryear{{Bernardi}, {Nichol}, {Sheth}, {Miller}  \&
  {Brinkmann}}{{Bernardi} et~al.}{2006}]{2006AJ....131.1288B}
{Bernardi} M.,  {Nichol} R.~C.,  {Sheth} R.~K.,  {Miller} C.~J.,   {Brinkmann}
  J.,  2006, \mn@doi [\aj] {10.1086/499522}, \href
  {http://adsabs.harvard.edu/abs/2006AJ....131.1288B} {131, 1288}

\bibitem[\protect\citeauthoryear{{Bottinelli}, {Gouguenheim}, {Paturel}  \&
  {Teerikorpi}}{{Bottinelli} et~al.}{1995}]{1995A&A...296...64B}
{Bottinelli} L.,  {Gouguenheim} L.,  {Paturel} G.,   {Teerikorpi} P.,  1995,
  \aap, \href {http://adsabs.harvard.edu/abs/1995A%26A...296...64B} {296, 64}

\bibitem[\protect\citeauthoryear{{Branch}, {Fisher}  \& {Nugent}}{{Branch}
  et~al.}{1993}]{1993AJ....106.2383B}
{Branch} D.,  {Fisher} A.,   {Nugent} P.,  1993, \mn@doi [\aj]
  {10.1086/116810}, \href {http://adsabs.harvard.edu/abs/1993AJ....106.2383B}
  {106, 2383}

\bibitem[\protect\citeauthoryear{{Capaccioli}}{{Capaccioli}}{1989}]{1989woga.conf..208C}
{Capaccioli} M.,  1989, in {Corwin} Jr. H.~G.,  {Bottinelli} L.,  eds, The
  World of Galaxies, New York: Springer-Verlag. p.~208

\bibitem[\protect\citeauthoryear{{Cappellaro} \& {Turatto}}{{Cappellaro} \&
  {Turatto}}{1997}]{1997ASIC..486...77C}
{Cappellaro} E.,  {Turatto} M.,  1997, in {Ruiz-Lapuente} P.,  {Canal} R.,
  {Isern} J.,  eds,  NATO Adv. Sci. Inst. Ser. C, Vol. 486, Thermonuclear
  Supernovae. Springer, Dordrecht. p.~77

\bibitem[\protect\citeauthoryear{{Chen} et~al.,}{{Chen}
  et~al.}{2012}]{2012MNRAS.421..314C}
{Chen} Y.-M.,  et~al., 2012, \mn@doi [\mnras]
  {10.1111/j.1365-2966.2011.20306.x}, \href
  {http://adsabs.harvard.edu/abs/2012MNRAS.421..314C} {421, 314}

\bibitem[\protect\citeauthoryear{{Childress}, {Wolf}  \& {Zahid}}{{Childress}
  et~al.}{2014}]{2014MNRAS.445.1898C}
{Childress} M.~J.,  {Wolf} C.,   {Zahid} H.~J.,  2014, \mn@doi [\mnras]
  {10.1093/mnras/stu1892}, \href
  {http://adsabs.harvard.edu/abs/2014MNRAS.445.1898C} {445, 1898}

\bibitem[\protect\citeauthoryear{{Conroy}, {Gunn}  \& {White}}{{Conroy}
  et~al.}{2009}]{2009ApJ...699..486C}
{Conroy} C.,  {Gunn} J.~E.,   {White} M.,  2009, \mn@doi [\apj]
  {10.1088/0004-637X/699/1/486}, \href
  {http://adsabs.harvard.edu/abs/2009ApJ...699..486C} {699, 486}

\bibitem[\protect\citeauthoryear{{Crocker} et~al.,}{{Crocker}
  et~al.}{2017}]{2017NatAs...1E.135C}
{Crocker} R.~M.,  et~al., 2017, \mn@doi [Nature Astronomy]
  {10.1038/s41550-017-0135}, \href
  {https://ui.adsabs.harvard.edu/abs/2017NatAs...1E.135C} {1, 0135}

\bibitem[\protect\citeauthoryear{{D'Agostino} \& {Stephens}}{{D'Agostino} \&
  {Stephens}}{1986}]{1986gft..book.....D}
{D'Agostino} R.~B.,  {Stephens} M.~A.,  1986, {Goodness-of-Fit Techniques,
  Statistics: Textbooks and Monographs, Vol.~68, Marcel Dekker Inc., New York,
  Basel}

\bibitem[\protect\citeauthoryear{{\SortNoop{De
  Vaucouleurs}}de~Vaucouleurs}{{\SortNoop{De
  Vaucouleurs}}de~Vaucouleurs}{1948}]{1948AnAp...11..247D}
{\SortNoop{De Vaucouleurs}}de~Vaucouleurs G.,  1948, Annales d'Astrophysique,
  \href {http://adsabs.harvard.edu/abs/1948AnAp...11..247D} {11, 247}

\bibitem[\protect\citeauthoryear{{Dilday} et~al.,}{{Dilday}
  et~al.}{2010}]{2010ApJ...715.1021D}
{Dilday} B.,  et~al., 2010, \mn@doi [\apj] {10.1088/0004-637X/715/2/1021},
  \href {http://adsabs.harvard.edu/abs/2010ApJ...715.1021D} {715, 1021}

\bibitem[\protect\citeauthoryear{{Dong}, {Katz}, {Kushnir}  \& {Prieto}}{{Dong}
  et~al.}{2015}]{2015MNRAS.454L..61D}
{Dong} S.,  {Katz} B.,  {Kushnir} D.,   {Prieto} J.~L.,  2015, \mn@doi [\mnras]
  {10.1093/mnrasl/slv129}, \href
  {http://adsabs.harvard.edu/abs/2015MNRAS.454L..61D} {454, L61}

\bibitem[\protect\citeauthoryear{{Engmann} \& {Cousineau}}{{Engmann} \&
  {Cousineau}}{2011}]{Engmann+11}
{Engmann} S.,  {Cousineau} D.,  2011, J. Appl. Quant. Methods, 6, 1

\bibitem[\protect\citeauthoryear{{Filippenko} et~al.,}{{Filippenko}
  et~al.}{1992a}]{1992AJ....104.1543F}
{Filippenko} A.~V.,  et~al., 1992a, \mn@doi [\aj] {10.1086/116339}, \href
  {http://adsabs.harvard.edu/abs/1992AJ....104.1543F} {104, 1543}

\bibitem[\protect\citeauthoryear{{Filippenko} et~al.,}{{Filippenko}
  et~al.}{1992b}]{1992ApJ...384L..15F}
{Filippenko} A.~V.,  et~al., 1992b, \mn@doi [\apjl] {10.1086/186252}, \href
  {http://adsabs.harvard.edu/abs/1992ApJ...384L..15F} {384, L15}

\bibitem[\protect\citeauthoryear{{Fioc} \& {Rocca-Volmerange}}{{Fioc} \&
  {Rocca-Volmerange}}{1997}]{1997A&A...326..950F}
{Fioc} M.,  {Rocca-Volmerange} B.,  1997, \aap, \href
  {http://adsabs.harvard.edu/abs/1997A%26A...326..950F} {326, 950}

\bibitem[\protect\citeauthoryear{{Fioc} \& {Rocca-Volmerange}}{{Fioc} \&
  {Rocca-Volmerange}}{1999}]{1999astro.ph.12179F}
{Fioc} M.,  {Rocca-Volmerange} B.,  1999, preprint, \href
  {http://adsabs.harvard.edu/abs/1999astro.ph.12179F} {} (\mn@eprint {}
  {astro-ph/9912179})

\bibitem[\protect\citeauthoryear{{F{\"o}rster} \& {Schawinski}}{{F{\"o}rster}
  \& {Schawinski}}{2008}]{2008MNRAS.388L..74F}
{F{\"o}rster} F.,  {Schawinski} K.,  2008, \mn@doi [\mnras]
  {10.1111/j.1745-3933.2008.00502.x}, \href
  {http://adsabs.harvard.edu/abs/2008MNRAS.388L..74F} {388, L74}

\bibitem[\protect\citeauthoryear{{Freeman}}{{Freeman}}{1970}]{1970ApJ...160..811F}
{Freeman} K.~C.,  1970, \mn@doi [\apj] {10.1086/150474}, \href
  {http://adsabs.harvard.edu/abs/1970ApJ...160..811F} {160, 811}

\bibitem[\protect\citeauthoryear{{Galbany} et~al.,}{{Galbany}
  et~al.}{2012}]{2012ApJ...755..125G}
{Galbany} L.,  et~al., 2012, \mn@doi [\apj] {10.1088/0004-637X/755/2/125},
  \href {http://adsabs.harvard.edu/abs/2012ApJ...755..125G} {755, 125}

\bibitem[\protect\citeauthoryear{{Gallagher}, {Garnavich}, {Berlind},
  {Challis}, {Jha}  \& {Kirshner}}{{Gallagher}
  et~al.}{2005}]{2005ApJ...634..210G}
{Gallagher} J.~S.,  {Garnavich} P.~M.,  {Berlind} P.,  {Challis} P.,  {Jha} S.,
    {Kirshner} R.~P.,  2005, \mn@doi [\apj] {10.1086/491664}, \href
  {http://adsabs.harvard.edu/abs/2005ApJ...634..210G} {634, 210}

\bibitem[\protect\citeauthoryear{{Gallagher}, {Garnavich}, {Caldwell},
  {Kirshner}, {Jha}, {Li}, {Ganeshalingam}  \& {Filippenko}}{{Gallagher}
  et~al.}{2008}]{2008ApJ...685..752G}
{Gallagher} J.~S.,  {Garnavich} P.~M.,  {Caldwell} N.,  {Kirshner} R.~P.,
  {Jha} S.~W.,  {Li} W.,  {Ganeshalingam} M.,   {Filippenko} A.~V.,  2008,
  \mn@doi [\apj] {10.1086/590659}, \href
  {http://adsabs.harvard.edu/abs/2008ApJ...685..752G} {685, 752}

\bibitem[\protect\citeauthoryear{{Gallazzi}, {Charlot}, {Brinchmann}  \&
  {White}}{{Gallazzi} et~al.}{2006}]{2006MNRAS.370.1106G}
{Gallazzi} A.,  {Charlot} S.,  {Brinchmann} J.,   {White} S.~D.~M.,  2006,
  \mn@doi [\mnras] {10.1111/j.1365-2966.2006.10548.x}, \href
  {http://adsabs.harvard.edu/abs/2006MNRAS.370.1106G} {370, 1106}

\bibitem[\protect\citeauthoryear{{Gilfanov} \& {Bogd{\'a}n}}{{Gilfanov} \&
  {Bogd{\'a}n}}{2010}]{2010Natur.463..924G}
{Gilfanov} M.,  {Bogd{\'a}n} {\'A}.,  2010, \mn@doi [\nat]
  {10.1038/nature08685}, \href
  {https://ui.adsabs.harvard.edu/abs/2010Natur.463..924G} {463, 924}

\bibitem[\protect\citeauthoryear{{Gomes} et~al.,}{{Gomes}
  et~al.}{2016a}]{2016A&A...585A..92G}
{Gomes} J.~M.,  et~al., 2016a, \mn@doi [\aap] {10.1051/0004-6361/201525974},
  \href {http://adsabs.harvard.edu/abs/2016A%26A...585A..92G} {585, A92}

\bibitem[\protect\citeauthoryear{{Gomes} et~al.,}{{Gomes}
  et~al.}{2016b}]{2016A&A...588A..68G}
{Gomes} J.~M.,  et~al., 2016b, \mn@doi [\aap] {10.1051/0004-6361/201525976},
  \href {http://adsabs.harvard.edu/abs/2016A%26A...588A..68G} {588, A68}

\bibitem[\protect\citeauthoryear{{Gonz{\'a}lez Delgado} et~al.,}{{Gonz{\'a}lez
  Delgado} et~al.}{2015}]{2015A&A...581A.103G}
{Gonz{\'a}lez Delgado} R.~M.,  et~al., 2015, \mn@doi [\aap]
  {10.1051/0004-6361/201525938}, \href
  {http://adsabs.harvard.edu/abs/2015A%26A...581A.103G} {581, A103}

\bibitem[\protect\citeauthoryear{{Gonz{\'a}lez-Gait{\'a}n}
  et~al.,}{{Gonz{\'a}lez-Gait{\'a}n} et~al.}{2011}]{2011ApJ...727..107G}
{Gonz{\'a}lez-Gait{\'a}n} S.,  et~al., 2011, \mn@doi [\apj]
  {10.1088/0004-637X/727/2/107}, \href
  {http://adsabs.harvard.edu/abs/2011ApJ...727..107G} {727, 107}

\bibitem[\protect\citeauthoryear{{Gonz{\'a}lez-Gait{\'a}n}
  et~al.,}{{Gonz{\'a}lez-Gait{\'a}n} et~al.}{2014}]{2014ApJ...795..142G}
{Gonz{\'a}lez-Gait{\'a}n} S.,  et~al., 2014, \mn@doi [\apj]
  {10.1088/0004-637X/795/2/142}, \href
  {http://adsabs.harvard.edu/abs/2014ApJ...795..142G} {795, 142}

\bibitem[\protect\citeauthoryear{{Guillochon}, {Parrent}, {Kelley}  \&
  {Margutti}}{{Guillochon} et~al.}{2017}]{2017ApJ...835...64G}
{Guillochon} J.,  {Parrent} J.,  {Kelley} L.~Z.,   {Margutti} R.,  2017,
  \mn@doi [\apj] {10.3847/1538-4357/835/1/64}, \href
  {http://adsabs.harvard.edu/abs/2017ApJ...835...64G} {835, 64}

\bibitem[\protect\citeauthoryear{{Gupta} et~al.,}{{Gupta}
  et~al.}{2011}]{2011ApJ...740...92G}
{Gupta} R.~R.,  et~al., 2011, \mn@doi [\apj] {10.1088/0004-637X/740/2/92},
  \href {http://adsabs.harvard.edu/abs/2011ApJ...740...92G} {740, 92}

\bibitem[\protect\citeauthoryear{{Guseinov}, {Kasumov}  \&
  {Kalinin}}{{Guseinov} et~al.}{1980}]{1980Ap&SS..68..385G}
{Guseinov} O.~H.,  {Kasumov} F.~K.,   {Kalinin} E.~V.,  1980, \mn@doi [\apss]
  {10.1007/BF00639706}, \href
  {http://adsabs.harvard.edu/abs/1980Ap%26SS..68..385G} {68, 385}

\bibitem[\protect\citeauthoryear{{Guy} et~al.,}{{Guy}
  et~al.}{2007}]{2007A&A...466...11G}
{Guy} J.,  et~al., 2007, \mn@doi [\aap] {10.1051/0004-6361:20066930}, \href
  {http://adsabs.harvard.edu/abs/2007A%26A...466...11G} {466, 11}

\bibitem[\protect\citeauthoryear{{Hakobyan} et~al.,}{{Hakobyan}
  et~al.}{2011}]{2011Ap.....54..301H}
{Hakobyan} A.~A.,  et~al., 2011, \mn@doi [Astrophysics]
  {10.1007/s10511-011-9180-y}, \href
  {http://adsabs.harvard.edu/abs/2011Ap.....54..301H} {54, 301}

\bibitem[\protect\citeauthoryear{{Hakobyan}, {Adibekyan}, {Aramyan},
  {Petrosian}, {Gomes}, {Mamon}, {Kunth}  \& {Turatto}}{{Hakobyan}
  et~al.}{2012}]{2012A&A...544A..81H}
{Hakobyan} A.~A.,  {Adibekyan} V.~Z.,  {Aramyan} L.~S.,  {Petrosian} A.~R.,
  {Gomes} J.~M.,  {Mamon} G.~A.,  {Kunth} D.,   {Turatto} M.,  2012, \mn@doi
  [\aap] {10.1051/0004-6361/201219541}, \href
  {http://adsabs.harvard.edu/abs/2012A%26A...544A..81H} {544, A81}

\bibitem[\protect\citeauthoryear{{Hakobyan} et~al.,}{{Hakobyan}
  et~al.}{2014}]{2014MNRAS.444.2428H}
{Hakobyan} A.~A.,  et~al., 2014, \mn@doi [\mnras] {10.1093/mnras/stu1598},
  \href {http://adsabs.harvard.edu/abs/2014MNRAS.444.2428H} {444, 2428}

\bibitem[\protect\citeauthoryear{{Hakobyan} et~al.,}{{Hakobyan}
  et~al.}{2016}]{2016MNRAS.456.2848H}
{Hakobyan} A.~A.,  et~al., 2016, \mn@doi [\mnras] {10.1093/mnras/stv2853},
  \href {http://adsabs.harvard.edu/abs/2016MNRAS.456.2848H} {456, 2848}

\bibitem[\protect\citeauthoryear{{Hakobyan} et~al.,}{{Hakobyan}
  et~al.}{2017}]{2017MNRAS.471.1390H}
{Hakobyan} A.~A.,  et~al., 2017, \mn@doi [\mnras] {10.1093/mnras/stx1608},
  \href {http://adsabs.harvard.edu/abs/2017MNRAS.471.1390H} {471, 1390}

\bibitem[\protect\citeauthoryear{{Hamuy} \& {Pinto}}{{Hamuy} \&
  {Pinto}}{1999}]{1999AJ....117.1185H}
{Hamuy} M.,  {Pinto} P.~A.,  1999, \mn@doi [\aj] {10.1086/300759}, \href
  {http://adsabs.harvard.edu/abs/1999AJ....117.1185H} {117, 1185}

\bibitem[\protect\citeauthoryear{{Hamuy}, {Phillips}, {Suntzeff}, {Schommer},
  {Maza}  \& {Aviles}}{{Hamuy} et~al.}{1996}]{1996AJ....112.2391H}
{Hamuy} M.,  {Phillips} M.~M.,  {Suntzeff} N.~B.,  {Schommer} R.~A.,  {Maza}
  J.,   {Aviles} R.,  1996, \mn@doi [\aj] {10.1086/118190}, \href
  {http://adsabs.harvard.edu/abs/1996AJ....112.2391H} {112, 2391}

\bibitem[\protect\citeauthoryear{{Hamuy}, {Trager}, {Pinto}, {Phillips},
  {Schommer}, {Ivanov}  \& {Suntzeff}}{{Hamuy}
  et~al.}{2000}]{2000AJ....120.1479H}
{Hamuy} M.,  {Trager} S.~C.,  {Pinto} P.~A.,  {Phillips} M.~M.,  {Schommer}
  R.~A.,  {Ivanov} V.,   {Suntzeff} N.~B.,  2000, \mn@doi [\aj]
  {10.1086/301527}, \href {http://adsabs.harvard.edu/abs/2000AJ....120.1479H}
  {120, 1479}

\bibitem[\protect\citeauthoryear{{Henry} \& {Worthey}}{{Henry} \&
  {Worthey}}{1999}]{1999PASP..111..919H}
{Henry} R.~B.~C.,  {Worthey} G.,  1999, \mn@doi [\pasp] {10.1086/316403}, \href
  {http://adsabs.harvard.edu/abs/1999PASP..111..919H} {111, 919}

\bibitem[\protect\citeauthoryear{{Hillebrandt} \& {Niemeyer}}{{Hillebrandt} \&
  {Niemeyer}}{2000}]{2000ARA&A..38..191H}
{Hillebrandt} W.,  {Niemeyer} J.~C.,  2000, \mn@doi [\araa]
  {10.1146/annurev.astro.38.1.191}, \href
  {http://adsabs.harvard.edu/abs/2000ARA%26A..38..191H} {38, 191}

\bibitem[\protect\citeauthoryear{{Howell}}{{Howell}}{2001}]{2001ApJ...554L.193H}
{Howell} D.~A.,  2001, \mn@doi [\apjl] {10.1086/321702}, \href
  {http://adsabs.harvard.edu/abs/2001ApJ...554L.193H} {554, L193}

\bibitem[\protect\citeauthoryear{{Howell}, {Wang}  \& {Wheeler}}{{Howell}
  et~al.}{2000}]{2000ApJ...530..166H}
{Howell} D.~A.,  {Wang} L.,   {Wheeler} J.~C.,  2000, \mn@doi [\apj]
  {10.1086/308356}, \href {http://adsabs.harvard.edu/abs/2000ApJ...530..166H}
  {530, 166}

\bibitem[\protect\citeauthoryear{{Howell} et~al.,}{{Howell}
  et~al.}{2009}]{2009ApJ...691..661H}
{Howell} D.~A.,  et~al., 2009, \mn@doi [\apj] {10.1088/0004-637X/691/1/661},
  \href {http://adsabs.harvard.edu/abs/2009ApJ...691..661H} {691, 661}

\bibitem[\protect\citeauthoryear{{Iben} \& {Tutukov}}{{Iben} \&
  {Tutukov}}{1984}]{1984ApJS...54..335I}
{Iben} Jr. I.,  {Tutukov} A.~V.,  1984, \mn@doi [\apjs] {10.1086/190932}, \href
  {http://adsabs.harvard.edu/abs/1984ApJS...54..335I} {54, 335}

\bibitem[\protect\citeauthoryear{{Ivanov}, {Hamuy}  \& {Pinto}}{{Ivanov}
  et~al.}{2000}]{2000ApJ...542..588I}
{Ivanov} V.~D.,  {Hamuy} M.,   {Pinto} P.~A.,  2000, \mn@doi [\apj]
  {10.1086/317060}, \href {http://adsabs.harvard.edu/abs/2000ApJ...542..588I}
  {542, 588}

\bibitem[\protect\citeauthoryear{{James} \& {Anderson}}{{James} \&
  {Anderson}}{2006}]{2006A&A...453...57J}
{James} P.~A.,  {Anderson} J.~P.,  2006, \mn@doi [\aap]
  {10.1051/0004-6361:20054509}, \href
  {http://adsabs.harvard.edu/abs/2006A%26A...453...57J} {453, 57}

\bibitem[\protect\citeauthoryear{{Jha}, {Riess}  \& {Kirshner}}{{Jha}
  et~al.}{2007}]{2007ApJ...659..122J}
{Jha} S.,  {Riess} A.~G.,   {Kirshner} R.~P.,  2007, \mn@doi [\apj]
  {10.1086/512054}, \href {http://adsabs.harvard.edu/abs/2007ApJ...659..122J}
  {659, 122}

\bibitem[\protect\citeauthoryear{{Kang}, {Kim}, {Lim}, {Chung}  \&
  {Lee}}{{Kang} et~al.}{2016}]{2016ApJS..223....7K}
{Kang} Y.,  {Kim} Y.-L.,  {Lim} D.,  {Chung} C.,   {Lee} Y.-W.,  2016, \mn@doi
  [\apjs] {10.3847/0067-0049/223/1/7}, \href
  {https://ui.adsabs.harvard.edu/abs/2016ApJS..223....7K} {223, 7}

\bibitem[\protect\citeauthoryear{{Kaviraj}, {Peirani}, {Khochfar}, {Silk}  \&
  {Kay}}{{Kaviraj} et~al.}{2009}]{2009MNRAS.394.1713K}
{Kaviraj} S.,  {Peirani} S.,  {Khochfar} S.,  {Silk} J.,   {Kay} S.,  2009,
  \mn@doi [\mnras] {10.1111/j.1365-2966.2009.14403.x}, \href
  {http://adsabs.harvard.edu/abs/2009MNRAS.394.1713K} {394, 1713}

\bibitem[\protect\citeauthoryear{{Kim}, {Smith}, {Sullivan}  \& {Lee}}{{Kim}
  et~al.}{2018}]{2018ApJ...854...24K}
{Kim} Y.-L.,  {Smith} M.,  {Sullivan} M.,   {Lee} Y.-W.,  2018, \mn@doi [\apj]
  {10.3847/1538-4357/aaa127}, \href
  {http://adsabs.harvard.edu/abs/2018ApJ...854...24K} {854, 24}

\bibitem[\protect\citeauthoryear{{Kochanek} et~al.,}{{Kochanek}
  et~al.}{2017}]{2017PASP..129j4502K}
{Kochanek} C.~S.,  et~al., 2017, \mn@doi [\pasp] {10.1088/1538-3873/aa80d9},
  \href {http://adsabs.harvard.edu/abs/2017PASP..129j4502K} {129, 104502}

\bibitem[\protect\citeauthoryear{{Kollmeier} et~al.,}{{Kollmeier}
  et~al.}{2019}]{2019MNRAS.486.3041K}
{Kollmeier} J.~A.,  et~al., 2019, \mn@doi [\mnras] {10.1093/mnras/stz953},
  \href {https://ui.adsabs.harvard.edu/abs/2019MNRAS.486.3041K} {486, 3041}

\bibitem[\protect\citeauthoryear{{Kormendy}, {Fisher}, {Cornell}  \&
  {Bender}}{{Kormendy} et~al.}{2009}]{2009ApJS..182..216K}
{Kormendy} J.,  {Fisher} D.~B.,  {Cornell} M.~E.,   {Bender} R.,  2009, \mn@doi
  [\apjs] {10.1088/0067-0049/182/1/216}, \href
  {http://adsabs.harvard.edu/abs/2009ApJS..182..216K} {182, 216}

\bibitem[\protect\citeauthoryear{{Leaman}, {Li}, {Chornock}  \&
  {Filippenko}}{{Leaman} et~al.}{2011}]{2011MNRAS.412.1419L}
{Leaman} J.,  {Li} W.,  {Chornock} R.,   {Filippenko} A.~V.,  2011, \mn@doi
  [\mnras] {10.1111/j.1365-2966.2011.18158.x}, \href
  {http://adsabs.harvard.edu/abs/2011MNRAS.412.1419L} {412, 1419}

\bibitem[\protect\citeauthoryear{{Leibundgut} et~al.,}{{Leibundgut}
  et~al.}{1993}]{1993AJ....105..301L}
{Leibundgut} B.,  et~al., 1993, \mn@doi [\aj] {10.1086/116427}, \href
  {https://ui.adsabs.harvard.edu/abs/1993AJ....105..301L} {105, 301}

\bibitem[\protect\citeauthoryear{{Li} et~al.,}{{Li}
  et~al.}{2011a}]{2011MNRAS.412.1441L}
{Li} W.,  et~al., 2011a, \mn@doi [\mnras] {10.1111/j.1365-2966.2011.18160.x},
  \href {http://adsabs.harvard.edu/abs/2011MNRAS.412.1441L} {412, 1441}

\bibitem[\protect\citeauthoryear{{Li}, {Chornock}, {Leaman}, {Filippenko},
  {Poznanski}, {Wang}, {Ganeshalingam}  \& {Mannucci}}{{Li}
  et~al.}{2011b}]{2011MNRAS.412.1473L}
{Li} W.,  {Chornock} R.,  {Leaman} J.,  {Filippenko} A.~V.,  {Poznanski} D.,
  {Wang} X.,  {Ganeshalingam} M.,   {Mannucci} F.,  2011b, \mn@doi [\mnras]
  {10.1111/j.1365-2966.2011.18162.x}, \href
  {http://adsabs.harvard.edu/abs/2011MNRAS.412.1473L} {412, 1473}

\bibitem[\protect\citeauthoryear{{Lipunov}, {Panchenko}  \&
  {Pruzhinskaya}}{{Lipunov} et~al.}{2011}]{2011NewA...16..250L}
{Lipunov} V.~M.,  {Panchenko} I.~E.,   {Pruzhinskaya} M.~V.,  2011, \mn@doi
  [\na] {10.1016/j.newast.2010.09.001}, \href
  {https://ui.adsabs.harvard.edu/abs/2011NewA...16..250L} {16, 250}

\bibitem[\protect\citeauthoryear{{Lupton}, {Gunn}, {Ivezi{\'c}}, {Knapp}  \&
  {Kent}}{{Lupton} et~al.}{2001}]{2001ASPC..238..269L}
{Lupton} R.,  {Gunn} J.~E.,  {Ivezi{\'c}} Z.,  {Knapp} G.~R.,   {Kent} S.,
  2001, in {Harnden} Jr. F.~R.,  {Primini} F.~A.,   {Payne} H.~E.,  eds,  ASP
  Conf. Ser. Vol. 238, Astronomical Data Analysis Software and Systems X.
  p.~269

\bibitem[\protect\citeauthoryear{{Maeda} \& {Terada}}{{Maeda} \&
  {Terada}}{2016}]{2016IJMPD..2530024M}
{Maeda} K.,  {Terada} Y.,  2016, \mn@doi [International Journal of Modern
  Physics D] {10.1142/S021827181630024X}, \href
  {http://adsabs.harvard.edu/abs/2016IJMPD..2530024M} {25, 1630024}

\bibitem[\protect\citeauthoryear{{Mannucci}, {Della Valle}, {Panagia},
  {Cappellaro}, {Cresci}, {Maiolino}, {Petrosian}  \& {Turatto}}{{Mannucci}
  et~al.}{2005}]{2005A&A...433..807M}
{Mannucci} F.,  {Della Valle} M.,  {Panagia} N.,  {Cappellaro} E.,  {Cresci}
  G.,  {Maiolino} R.,  {Petrosian} A.,   {Turatto} M.,  2005, \mn@doi [\aap]
  {10.1051/0004-6361:20041411}, \href
  {http://adsabs.harvard.edu/abs/2005A%26A...433..807M} {433, 807}

\bibitem[\protect\citeauthoryear{{Maoz}, {Mannucci}  \& {Nelemans}}{{Maoz}
  et~al.}{2014}]{2014ARA&A..52..107M}
{Maoz} D.,  {Mannucci} F.,   {Nelemans} G.,  2014, \mn@doi [\araa]
  {10.1146/annurev-astro-082812-141031}, \href
  {http://adsabs.harvard.edu/abs/2014ARA%26A..52..107M} {52, 107}

\bibitem[\protect\citeauthoryear{{Maraston}}{{Maraston}}{2005}]{2005MNRAS.362..799M}
{Maraston} C.,  2005, \mn@doi [\mnras] {10.1111/j.1365-2966.2005.09270.x},
  \href {http://adsabs.harvard.edu/abs/2005MNRAS.362..799M} {362, 799}

\bibitem[\protect\citeauthoryear{{Massey}}{{Massey}}{1951}]{Massey51}
{Massey} F.~J.,  1951, J. Am. Stat. Assoc., 46, 68

\bibitem[\protect\citeauthoryear{{Maza} \& {van den Bergh}}{{Maza} \& {van den
  Bergh}}{1976}]{1976ApJ...204..519M}
{Maza} J.,  {van den Bergh} S.,  1976, \mn@doi [\apj] {10.1086/154198}, \href
  {http://adsabs.harvard.edu/abs/1976ApJ...204..519M} {204, 519}

\bibitem[\protect\citeauthoryear{{Mazzali} \& {Hachinger}}{{Mazzali} \&
  {Hachinger}}{2012}]{2012MNRAS.424.2926M}
{Mazzali} P.~A.,  {Hachinger} S.,  2012, \mn@doi [\mnras]
  {10.1111/j.1365-2966.2012.21433.x}, \href
  {http://adsabs.harvard.edu/abs/2012MNRAS.424.2926M} {424, 2926}

\bibitem[\protect\citeauthoryear{{Mazzali}, {R{\"o}pke}, {Benetti}  \&
  {Hillebrandt}}{{Mazzali} et~al.}{2007}]{2007Sci...315..825M}
{Mazzali} P.~A.,  {R{\"o}pke} F.~K.,  {Benetti} S.,   {Hillebrandt} W.,  2007,
  \mn@doi [Science] {10.1126/science.1136259}, \href
  {http://adsabs.harvard.edu/abs/2007Sci...315..825M} {315, 825}

\bibitem[\protect\citeauthoryear{{McIntosh} et~al.,}{{McIntosh}
  et~al.}{2014}]{2014MNRAS.442..533M}
{McIntosh} D.~H.,  et~al., 2014, \mn@doi [\mnras] {10.1093/mnras/stu808}, \href
  {http://adsabs.harvard.edu/abs/2014MNRAS.442..533M} {442, 533}

\bibitem[\protect\citeauthoryear{{Mendez}, {Coil}, {Lotz}, {Salim}, {Moustakas}
   \& {Simard}}{{Mendez} et~al.}{2011}]{2011ApJ...736..110M}
{Mendez} A.~J.,  {Coil} A.~L.,  {Lotz} J.,  {Salim} S.,  {Moustakas} J.,
  {Simard} L.,  2011, \mn@doi [\apj] {10.1088/0004-637X/736/2/110}, \href
  {http://adsabs.harvard.edu/abs/2011ApJ...736..110M} {736, 110}

\bibitem[\protect\citeauthoryear{{Moreno-Raya}, {L{\'o}pez-S{\'a}nchez},
  {Moll{\'a}}, {Galbany}, {V{\'{\i}}lchez}  \& {Carnero}}{{Moreno-Raya}
  et~al.}{2016}]{2016MNRAS.462.1281M}
{Moreno-Raya} M.~E.,  {L{\'o}pez-S{\'a}nchez} {\'A}.~R.,  {Moll{\'a}} M.,
  {Galbany} L.,  {V{\'{\i}}lchez} J.~M.,   {Carnero} A.,  2016, \mn@doi
  [\mnras] {10.1093/mnras/stw1706}, \href
  {https://ui.adsabs.harvard.edu/abs/2016MNRAS.462.1281M} {462, 1281}

\bibitem[\protect\citeauthoryear{{Nair} \& {Abraham}}{{Nair} \&
  {Abraham}}{2010}]{2010ApJS..186..427N}
{Nair} P.~B.,  {Abraham} R.~G.,  2010, \mn@doi [\apjs]
  {10.1088/0067-0049/186/2/427}, \href
  {http://adsabs.harvard.edu/abs/2010ApJS..186..427N} {186, 427}

\bibitem[\protect\citeauthoryear{{Neill} et~al.,}{{Neill}
  et~al.}{2009}]{2009ApJ...707.1449N}
{Neill} J.~D.,  et~al., 2009, \mn@doi [\apj] {10.1088/0004-637X/707/2/1449},
  \href {http://adsabs.harvard.edu/abs/2009ApJ...707.1449N} {707, 1449}

\bibitem[\protect\citeauthoryear{{Nomoto}, {Iwamoto}  \& {Kishimoto}}{{Nomoto}
  et~al.}{1997}]{1997Sci...276.1378N}
{Nomoto} K.,  {Iwamoto} K.,   {Kishimoto} N.,  1997, \mn@doi [Science]
  {10.1126/science.276.5317.1378}, \href
  {http://adsabs.harvard.edu/abs/1997Sci...276.1378N} {276, 1378}

\bibitem[\protect\citeauthoryear{{Pakmor}, {Kromer}, {Taubenberger}  \&
  {Springel}}{{Pakmor} et~al.}{2013}]{2013ApJ...770L...8P}
{Pakmor} R.,  {Kromer} M.,  {Taubenberger} S.,   {Springel} V.,  2013, \mn@doi
  [\apjl] {10.1088/2041-8205/770/1/L8}, \href
  {https://ui.adsabs.harvard.edu/abs/2013ApJ...770L...8P} {770, L8}

\bibitem[\protect\citeauthoryear{{Pan} et~al.,}{{Pan}
  et~al.}{2014}]{2014MNRAS.438.1391P}
{Pan} Y.-C.,  et~al., 2014, \mn@doi [\mnras] {10.1093/mnras/stt2287}, \href
  {http://adsabs.harvard.edu/abs/2014MNRAS.438.1391P} {438, 1391}

\bibitem[\protect\citeauthoryear{{Pan}, {Sullivan}, {Maguire}, {Gal-Yam},
  {Hook}, {Howell}, {Nugent}  \& {Mazzali}}{{Pan}
  et~al.}{2015}]{2015MNRAS.446..354P}
{Pan} Y.-C.,  {Sullivan} M.,  {Maguire} K.,  {Gal-Yam} A.,  {Hook} I.~M.,
  {Howell} D.~A.,  {Nugent} P.~E.,   {Mazzali} P.~A.,  2015, \mn@doi [\mnras]
  {10.1093/mnras/stu2121}, \href
  {http://adsabs.harvard.edu/abs/2015MNRAS.446..354P} {446, 354}

\bibitem[\protect\citeauthoryear{{Panther}, {Seitenzahl}, {Ruiter}, {Crocker},
  {Lidman}, {Wang}, {Tucker}  \& {Groves}}{{Panther}
  et~al.}{2019}]{2019arXiv190410139P}
{Panther} F.~H.,  {Seitenzahl} I.~R.,  {Ruiter} A.~J.,  {Crocker} R.~M.,
  {Lidman} C.,  {Wang} E.~X.,  {Tucker} B.~E.,   {Groves} B.,  2019, preprint,
  \href {http://adsabs.harvard.edu/abs/2019arXiv190410139P} {} (\mn@eprint
  {arXiv} {1904.10139})

\bibitem[\protect\citeauthoryear{{Pavlyuk} \& {Tsvetkov}}{{Pavlyuk} \&
  {Tsvetkov}}{2016}]{2016AstL...42..495P}
{Pavlyuk} N.~N.,  {Tsvetkov} D.~Y.,  2016, \mn@doi [Astron. Lett.]
  {10.1134/S1063773716080053}, \href
  {http://adsabs.harvard.edu/abs/2016AstL...42..495P} {42, 495}

\bibitem[\protect\citeauthoryear{{Perlmutter} et~al.,}{{Perlmutter}
  et~al.}{1999}]{1999ApJ...517..565P}
{Perlmutter} S.,  et~al., 1999, \mn@doi [\apj] {10.1086/307221}, \href
  {http://adsabs.harvard.edu/abs/1999ApJ...517..565P} {517, 565}

\bibitem[\protect\citeauthoryear{{Pettitt}}{{Pettitt}}{1976}]{Pettitt76}
{Pettitt} A.~N.,  1976, Biometrika, 63, 161

\bibitem[\protect\citeauthoryear{{Phillips}}{{Phillips}}{1993}]{1993ApJ...413L.105P}
{Phillips} M.~M.,  1993, \mn@doi [\apjl] {10.1086/186970}, \href
  {http://adsabs.harvard.edu/abs/1993ApJ...413L.105P} {413, L105}

\bibitem[\protect\citeauthoryear{{Phillips}, {Wells}, {Suntzeff}, {Hamuy},
  {Leibundgut}, {Kirshner}  \& {Foltz}}{{Phillips}
  et~al.}{1992}]{1992AJ....103.1632P}
{Phillips} M.~M.,  {Wells} L.~A.,  {Suntzeff} N.~B.,  {Hamuy} M.,  {Leibundgut}
  B.,  {Kirshner} R.~P.,   {Foltz} C.~B.,  1992, \mn@doi [\aj]
  {10.1086/116177}, \href
  {https://ui.adsabs.harvard.edu/abs/1992AJ....103.1632P} {103, 1632}

\bibitem[\protect\citeauthoryear{{Phillips}, {Lira}, {Suntzeff}, {Schommer},
  {Hamuy}  \& {Maza}}{{Phillips} et~al.}{1999}]{1999AJ....118.1766P}
{Phillips} M.~M.,  {Lira} P.,  {Suntzeff} N.~B.,  {Schommer} R.~A.,  {Hamuy}
  M.,   {Maza} J.,  1999, \mn@doi [\aj] {10.1086/301032}, \href
  {http://adsabs.harvard.edu/abs/1999AJ....118.1766P} {118, 1766}

\bibitem[\protect\citeauthoryear{{Pskovskii}}{{Pskovskii}}{1977}]{1977SvA....21..675P}
{Pskovskii} I.~P.,  1977, \sovast, \href
  {https://ui.adsabs.harvard.edu/abs/1977SvA....21..675P} {21, 675}

\bibitem[\protect\citeauthoryear{{Riess} et~al.,}{{Riess}
  et~al.}{1998}]{1998AJ....116.1009R}
{Riess} A.~G.,  et~al., 1998, \mn@doi [\aj] {10.1086/300499}, \href
  {http://adsabs.harvard.edu/abs/1998AJ....116.1009R} {116, 1009}

\bibitem[\protect\citeauthoryear{{Rose}, {Garnavich}  \& {Berg}}{{Rose}
  et~al.}{2019}]{2019ApJ...874...32R}
{Rose} B.~M.,  {Garnavich} P.~M.,   {Berg} M.~A.,  2019, \mn@doi [\apj]
  {10.3847/1538-4357/ab0704}, \href
  {https://ui.adsabs.harvard.edu/abs/2019ApJ...874...32R} {874, 32}

\bibitem[\protect\citeauthoryear{{Ruiz-Lapuente}, {Cappellaro}, {Turatto},
  {Gouiffes}, {Danziger}, {della Valle}  \& {Lucy}}{{Ruiz-Lapuente}
  et~al.}{1992}]{1992ApJ...387L..33R}
{Ruiz-Lapuente} P.,  {Cappellaro} E.,  {Turatto} M.,  {Gouiffes} C.,
  {Danziger} I.~J.,  {della Valle} M.,   {Lucy} L.~B.,  1992, \mn@doi [\apjl]
  {10.1086/186299}, \href
  {https://ui.adsabs.harvard.edu/abs/1992ApJ...387L..33R} {387, L33}

\bibitem[\protect\citeauthoryear{{Rust}}{{Rust}}{1974}]{1974PhDT.........7R}
{Rust} B.~W.,  1974, PhD thesis, Oak Ridge National Laboratory

\bibitem[\protect\citeauthoryear{{Sand} et~al.,}{{Sand}
  et~al.}{2019}]{2019ApJ...877L...4S}
{Sand} D.~J.,  et~al., 2019, \mn@doi [\apjl] {10.3847/2041-8213/ab1eaf}, \href
  {https://ui.adsabs.harvard.edu/abs/2019ApJ...877L...4S} {877, L4}

\bibitem[\protect\citeauthoryear{{Scannapieco} \& {Bildsten}}{{Scannapieco} \&
  {Bildsten}}{2005}]{2005ApJ...629L..85S}
{Scannapieco} E.,  {Bildsten} L.,  2005, \mn@doi [\apjl] {10.1086/452632},
  \href {http://adsabs.harvard.edu/abs/2005ApJ...629L..85S} {629, L85}

\bibitem[\protect\citeauthoryear{{Schawinski}, {Dowlin}, {Thomas}, {Urry}  \&
  {Edmondson}}{{Schawinski} et~al.}{2010}]{2010ApJ...714L.108S}
{Schawinski} K.,  {Dowlin} N.,  {Thomas} D.,  {Urry} C.~M.,   {Edmondson} E.,
  2010, \mn@doi [\apjl] {10.1088/2041-8205/714/1/L108}, \href
  {http://adsabs.harvard.edu/abs/2010ApJ...714L.108S} {714, L108}

\bibitem[\protect\citeauthoryear{{Schawinski} et~al.,}{{Schawinski}
  et~al.}{2014}]{2014MNRAS.440..889S}
{Schawinski} K.,  et~al., 2014, \mn@doi [\mnras] {10.1093/mnras/stu327}, \href
  {http://adsabs.harvard.edu/abs/2014MNRAS.440..889S} {440, 889}

\bibitem[\protect\citeauthoryear{{Schlafly} \& {Finkbeiner}}{{Schlafly} \&
  {Finkbeiner}}{2011}]{2011ApJ...737..103S}
{Schlafly} E.~F.,  {Finkbeiner} D.~P.,  2011, \mn@doi [\apj]
  {10.1088/0004-637X/737/2/103}, \href
  {http://adsabs.harvard.edu/abs/2011ApJ...737..103S} {737, 103}

\bibitem[\protect\citeauthoryear{{Schlegel}, {Finkbeiner}  \&
  {Davis}}{{Schlegel} et~al.}{1998}]{1998ApJ...500..525S}
{Schlegel} D.~J.,  {Finkbeiner} D.~P.,   {Davis} M.,  1998, \mn@doi [\apj]
  {10.1086/305772}, \href {http://adsabs.harvard.edu/abs/1998ApJ...500..525S}
  {500, 525}

\bibitem[\protect\citeauthoryear{{Scott} et~al.,}{{Scott}
  et~al.}{2017}]{2017MNRAS.472.2833S}
{Scott} N.,  et~al., 2017, \mn@doi [\mnras] {10.1093/mnras/stx2166}, \href
  {http://adsabs.harvard.edu/abs/2017MNRAS.472.2833S} {472, 2833}

\bibitem[\protect\citeauthoryear{{S{\'e}rsic}}{{S{\'e}rsic}}{1963}]{1963BAAA....6...41S}
{S{\'e}rsic} J.~L.,  1963, BAAA, \href
  {http://adsabs.harvard.edu/abs/1963BAAA....6...41S} {6, 41}

\bibitem[\protect\citeauthoryear{{Shaw}}{{Shaw}}{1979}]{1979A&A....76..188S}
{Shaw} R.~L.,  1979, \aap, \href
  {http://adsabs.harvard.edu/abs/1979A%26A....76..188S} {76, 188}

\bibitem[\protect\citeauthoryear{{Silverman} et~al.,}{{Silverman}
  et~al.}{2012}]{2012MNRAS.425.1789S}
{Silverman} J.~M.,  et~al., 2012, \mn@doi [\mnras]
  {10.1111/j.1365-2966.2012.21270.x}, \href
  {http://adsabs.harvard.edu/abs/2012MNRAS.425.1789S} {425, 1789}

\bibitem[\protect\citeauthoryear{{Sullivan} et~al.,}{{Sullivan}
  et~al.}{2010}]{2010MNRAS.406..782S}
{Sullivan} M.,  et~al., 2010, \mn@doi [\mnras]
  {10.1111/j.1365-2966.2010.16731.x}, \href
  {http://adsabs.harvard.edu/abs/2010MNRAS.406..782S} {406, 782}

\bibitem[\protect\citeauthoryear{{Taubenberger}}{{Taubenberger}}{2017}]{2017hsn..book..317T}
{Taubenberger} S.,  2017, {The Extremes of Thermonuclear Supernovae. In:
  Alsabti~A.~W., Murdin~P. (eds) Handbook of Supernovae. Springer, Cham}.
p.~317

\bibitem[\protect\citeauthoryear{{Taubenberger} et~al.,}{{Taubenberger}
  et~al.}{2008}]{2008MNRAS.385...75T}
{Taubenberger} S.,  et~al., 2008, \mn@doi [\mnras]
  {10.1111/j.1365-2966.2008.12843.x}, \href
  {http://adsabs.harvard.edu/abs/2008MNRAS.385...75T} {385, 75}

\bibitem[\protect\citeauthoryear{{Taubenberger} et~al.,}{{Taubenberger}
  et~al.}{2011}]{2011MNRAS.412.2735T}
{Taubenberger} S.,  et~al., 2011, \mn@doi [\mnras]
  {10.1111/j.1365-2966.2010.18107.x}, \href
  {https://ui.adsabs.harvard.edu/abs/2011MNRAS.412.2735T} {412, 2735}

\bibitem[\protect\citeauthoryear{{Taylor} et~al.,}{{Taylor}
  et~al.}{2011}]{2011MNRAS.418.1587T}
{Taylor} E.~N.,  et~al., 2011, \mn@doi [\mnras]
  {10.1111/j.1365-2966.2011.19536.x}, \href
  {http://adsabs.harvard.edu/abs/2011MNRAS.418.1587T} {418, 1587}

\bibitem[\protect\citeauthoryear{{Terry}, {Paturel}  \& {Ekholm}}{{Terry}
  et~al.}{2002}]{2002A&A...393...57T}
{Terry} J.~N.,  {Paturel} G.,   {Ekholm} T.,  2002, \mn@doi [\aap]
  {10.1051/0004-6361:20021018}, \href
  {http://adsabs.harvard.edu/abs/2002A%26A...393...57T} {393, 57}

\bibitem[\protect\citeauthoryear{{Theureau}, {Rauzy}, {Bottinelli}  \&
  {Gouguenheim}}{{Theureau} et~al.}{1998}]{1998A&A...340...21T}
{Theureau} G.,  {Rauzy} S.,  {Bottinelli} L.,   {Gouguenheim} L.,  1998, \aap,
  \href {http://adsabs.harvard.edu/abs/1998A%26A...340...21T} {340, 21}

\bibitem[\protect\citeauthoryear{{Timmes}, {Brown}  \& {Truran}}{{Timmes}
  et~al.}{2003}]{2003ApJ...590L..83T}
{Timmes} F.~X.,  {Brown} E.~F.,   {Truran} J.~W.,  2003, \mn@doi [\apjl]
  {10.1086/376721}, \href {http://adsabs.harvard.edu/abs/2003ApJ...590L..83T}
  {590, L83}

\bibitem[\protect\citeauthoryear{{Tomasella } et~al.,}{{Tomasella }
  et~al.}{2014}]{2014AN....335..841T}
{Tomasella } L.,  et~al., 2014, \mn@doi [Astronomische Nachrichten]
  {10.1002/asna.201412068}, \href
  {http://adsabs.harvard.edu/abs/2014AN....335..841T} {335, 841}

\bibitem[\protect\citeauthoryear{{Tsvetkov}, {Pavlyuk}  \&
  {Bartunov}}{{Tsvetkov} et~al.}{2004}]{2004AstL...30..729T}
{Tsvetkov} D.~Y.,  {Pavlyuk} N.~N.,   {Bartunov} O.~S.,  2004, \mn@doi [Astron.
  Lett.] {10.1134/1.1819491}, \href
  {http://adsabs.harvard.edu/abs/2004AstL...30..729T} {30, 729}

\bibitem[\protect\citeauthoryear{{Turatto}, {Benetti}, {Cappellaro},
  {Danziger}, {Della Valle}, {Gouiffes}, {Mazzali}  \& {Patat}}{{Turatto}
  et~al.}{1996}]{1996MNRAS.283....1T}
{Turatto} M.,  {Benetti} S.,  {Cappellaro} E.,  {Danziger} I.~J.,  {Della
  Valle} M.,  {Gouiffes} C.,  {Mazzali} P.~A.,   {Patat} F.,  1996, \mn@doi
  [\mnras] {10.1093/mnras/283.1.1}, \href
  {https://ui.adsabs.harvard.edu/abs/1996MNRAS.283....1T} {283, 1}

\bibitem[\protect\citeauthoryear{{Tyson}}{{Tyson}}{2002}]{2002SPIE.4836...10T}
{Tyson} J.~A.,  2002, in {Tyson} J.~A.,  {Wolff} S.,  eds,  \procspie Vol.
  4836, Survey and Other Telescope Technologies and Discoveries. pp 10--20

\bibitem[\protect\citeauthoryear{{Uddin}, {Mould}  \& {Wang}}{{Uddin}
  et~al.}{2017}]{2017ApJ...850..135U}
{Uddin} S.~A.,  {Mould} J.,   {Wang} L.,  2017, \mn@doi [\apj]
  {10.3847/1538-4357/aa93e9}, \href
  {https://ui.adsabs.harvard.edu/abs/2017ApJ...850..135U} {850, 135}

\bibitem[\protect\citeauthoryear{{Verkhodanov}, {Kopylov}, {Zhelenkova},
  {Verkhodanova}, {Chernenkov}, {Parijskij}, {Soboleva}  \&
  {Temirova}}{{Verkhodanov} et~al.}{2000}]{2000A&AT...19..662V}
{Verkhodanov} O.~V.,  {Kopylov} A.~I.,  {Zhelenkova} O.~P.,  {Verkhodanova}
  N.~V.,  {Chernenkov} V.~N.,  {Parijskij} Y.~N.,  {Soboleva} N.~S.,
  {Temirova} A.~V.,  2000, Astron. Astrophys. Trans., \href
  {http://adsabs.harvard.edu/abs/2000A%26AT...19..662V} {19, 662}

\bibitem[\protect\citeauthoryear{{Vika}, {Bamford}, {H{\"a}u{\ss}ler}, {Rojas},
  {Borch}  \& {Nichol}}{{Vika} et~al.}{2013}]{2013MNRAS.435..623V}
{Vika} M.,  {Bamford} S.~P.,  {H{\"a}u{\ss}ler} B.,  {Rojas} A.~L.,  {Borch}
  A.,   {Nichol} R.~C.,  2013, \mn@doi [\mnras] {10.1093/mnras/stt1320}, \href
  {http://adsabs.harvard.edu/abs/2013MNRAS.435..623V} {435, 623}

\bibitem[\protect\citeauthoryear{{Wang}, {Wang}, {Filippenko}, {Zhang}  \&
  {Zhao}}{{Wang} et~al.}{2013}]{2013Sci...340..170W}
{Wang} X.,  {Wang} L.,  {Filippenko} A.~V.,  {Zhang} T.,   {Zhao} X.,  2013,
  \mn@doi [Science] {10.1126/science.1231502}, \href
  {http://adsabs.harvard.edu/abs/2013Sci...340..170W} {340, 170}

\bibitem[\protect\citeauthoryear{{Yahil}, {Tammann}  \& {Sandage}}{{Yahil}
  et~al.}{1977}]{1977ApJ...217..903Y}
{Yahil} A.,  {Tammann} G.~A.,   {Sandage} A.,  1977, \mn@doi [\apj]
  {10.1086/155636}, \href {http://adsabs.harvard.edu/abs/1977ApJ...217..903Y}
  {217, 903}

\bibitem[\protect\citeauthoryear{{Yaron} \& {Gal-Yam}}{{Yaron} \&
  {Gal-Yam}}{2012}]{2012PASP..124..668Y}
{Yaron} O.,  {Gal-Yam} A.,  2012, \mn@doi [\pasp] {10.1086/666656}, \href
  {http://adsabs.harvard.edu/abs/2012PASP..124..668Y} {124, 668}

\makeatother
\end{thebibliography}

\section*{Supporting information}

Supplementary data are available at \emph{MNRAS} online.\\
\\
\textbf{PaperVIonlinedata.csv}
\\
\\
Please note: Oxford University Press is not responsible for the
content or functionality of any supporting materials supplied by
the authors. Any queries (other than missing material) should be
directed to the corresponding author for the article.

\label{lastpage}

\end{document}